\documentclass[aps,notitlepage,groupedaddress,longbibliography]{revtex4-1}

\pdfoutput=1

\usepackage{amsmath}
\usepackage{amssymb,amsfonts}
\usepackage{graphicx}
\usepackage{bm}
\usepackage{subfigure}
\usepackage{color}
\usepackage{hyperref}
\usepackage{wasysym}

\graphicspath{{figs/}}

\newcommand{\mathnotation}[2]{\newcommand{#1}{\ensuremath{#2}}}

\newcommand{\nofrac}[2]{#1/#2}

\DeclareMathOperator{\erfc}{erfc}
\DeclareMathOperator{\sinc}{sinc}

\renewcommand{\l}{\left}
\let\rsave\r 
\renewcommand{\r}{\right}
\mathnotation{\pd}{\partial}
\mathnotation{\ldef}{\mathrel{\raisebox{.069ex}{:}\!\!=}}
\mathnotation{\rdef}{\mathrel{=\!\!\raisebox{.069ex}{:}}}
\mathnotation{\dint}{\,{\mathrm{d}}}        
\mathnotation{\ee}{\mathrm{e}}              
\mathnotation{\imi}{\mathrm{i}}             
\mathnotation{\Wc}{W}                       
\renewcommand{\time}{t}                     
\mathnotation{\Impv}{\bm{I}}                

\newcommand{\Rey}{\mathrm{Re}}              

\mathnotation{\uv}{\bm{u}}
\mathnotation{\vv}{\bm{v}}
\mathnotation{\uc}{u}
\mathnotation{\vc}{v}
\mathnotation{\wc}{w}
\mathnotation{\rv}{\bm{r}}
\mathnotation{\xc}{x}
\mathnotation{\yc}{y}
\mathnotation{\zc}{z}
\mathnotation{\zuv}{\hat{\bm{z}}}
\mathnotation{\prob}{p}
\mathnotation{\ucmin}{\varepsilon}

\mathnotation{\grad}{\nabla}
\mathnotation{\lapl}{\grad^2}
\mathnotation{\pressure}{p}
\mathnotation{\force}{\bm{F}}

\mathnotation{\numV}{N}

\mathnotation{\T}{T}

\mathnotation{\Ss}{S}
\mathnotation{\U}{\bm{U}}
\mathnotation{\Rot}{\mathbb{Q}}
\mathnotation{\Rv}{\bm{R}}

\mathnotation{\prodrate}{\mu}
\mathnotation{\volfrac}{\phi}
\mathnotation{\hyperParam}{a}
\mathnotation{\hyperArg}{b}
\mathnotation{\cutoff}{c}
\mathnotation{\azAng}{\varphi}
\mathnotation{\angFunc}{f(\azAng)}
\mathnotation{\dAngFunc}{f^\prime(\azAng)}

\begin{document}

\title{Velocity fluctuations in a dilute suspension of viscous vortex rings}

\author{Thomas Morrell}
\email[]{tamorrell@math.wisc.edu}
\author{Saverio Spagnolie}
\email[]{spagnolie@math.wisc.edu}
\author{Jean-Luc Thiffeault}
\email[]{jeanluc@math.wisc.edu}
\affiliation{Department of Mathematics, University of Wisconsin--Madison, 480 Lincoln Drive, Madison, Wisconsin 53706, USA}

\date{\today}

\begin{abstract}
	We explore the velocity fluctuations in a fluid due to a dilute suspension of randomly-distributed vortex rings at moderate Reynolds number, for instance those generated by a large colony of jellyfish.  Unlike previous analysis of velocity fluctuations associated with gravitational sedimentation or suspensions of microswimmers, here the vortices have a finite lifetime and are constantly being produced. We find that the net velocity distribution is similar to that of a single vortex, except for the smallest velocities which involve contributions from many distant vortices; the result is a truncated $5/3$-stable distribution with variance (and mean energy) linear in the vortex volume fraction $\volfrac$. The distribution has an inner core with a width scaling as $\volfrac^{3/5}$, then long tails with power law $\lvert\uc\rvert^{-8/3}$, and finally a fixed cutoff (independent of $\volfrac$) above which the probability density scales as $\lvert\uc\rvert^{-5}$, where $\uc$ is a component of the velocity. We argue that this distribution is robust in the sense that the distribution of any velocity fluctuations caused by random forces localized in space and time has the same properties, except possibly for a different scaling after the cutoff.
\end{abstract}

\pacs{47}

\maketitle

\section{Introduction}
\label{Intro}

A natural question when faced with a fluid flow with some degree of randomness is how to characterize its velocity fluctuations.  This is a classical problem in turbulence, but also in gravitational sedimentation \cite{caflisch1985,Nicolai1995,luke2000,mucha2004,Guazzelli2011,moller2017}, and in suspensions of microswimmers \cite{Dombrowski2004,Drescher2010,Guasto2010,Ishikawa2007b,Ishikawa2009,Leptos2009,Rushkin2010,Underhill2008,Yeomans2014,Lin2011,Zaid2011,Delmotte2017}.  In the case of sedimentation and microswimmers, the velocity field due to a single particle or swimmer is commonly used as a building block to understand the velocity distribution in the full system. At leading order for a dilute suspension, interactions are neglected and much is learned by examining a random superposition of individual particles or swimmers.  In particular, for small velocities the distribution is typically Gaussian~\cite{Delmotte2017}, since superimposing many distant sources results in an application of the central limit theorem.

\begin{figure}
	\centering
	\includegraphics[width=\textwidth]{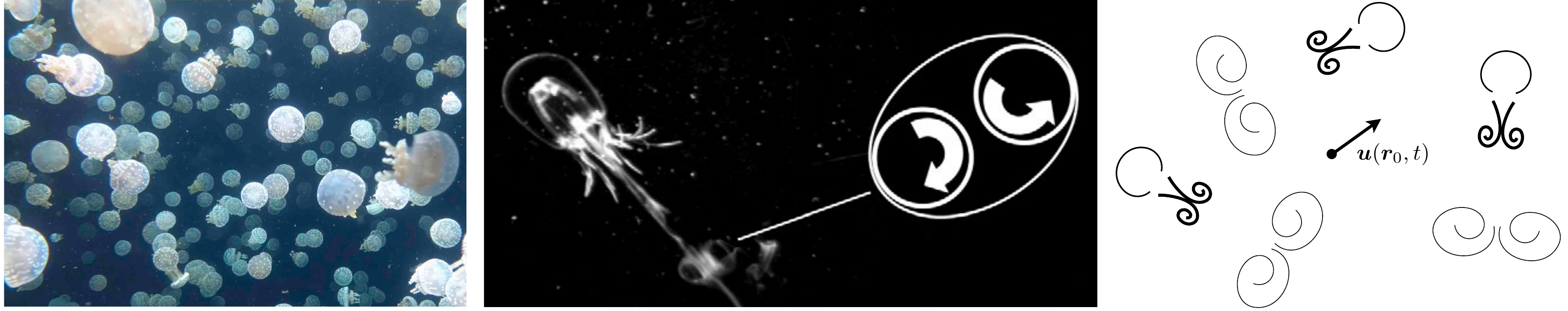}
	\caption{(Left) A ``suspension'' of spotted jellyfish ({\it Mastigias papua}) at the Vancouver Aquarium. (Center) Fast swimming {\it  Nemopsis bachei} expels a single vortex ring with each rapid pulse (reproduced with permission from \cite{Costello2006}). (Right) Schematic of the problem: we seek the distribution of the fluid velocity $\uv$ at $\rv_0$ due to a randomly distributed suspension of viscous vortex rings in three dimensions.}
	\label{fig:Jellies}
\end{figure}

In this paper we study the velocity distribution in a dilute suspension of viscous vortex rings. We assume some mechanism, such as a colony of jellyfish, generates vortices randomly throughout time and space, as observed and illustrated in Figure~\ref{fig:Jellies}. These vortices decay due to viscosity but are replenished such that the system is assumed to reach a statistical equilibrium, containing vortices with some age distribution.  Turbulence has been modeled with some success using vortex rings~\cite{Synge1943,phillips_final_1956,saffman_vortex_1997}, but here we investigate a moderate Reynolds number regime which is still a long way from turbulence (the jellyfish are assumed to be a few centimeters in size so that the rings they generate are strongly affected by viscosity). Other related biological systems may also exhibit related velocity field fluctuations that may have important functional consequences. In particular, non-motile pulsing corals share considerable hydrodynamic similarities with undulating jellyfish, and their repeated pulsing is known to contribute to fluid mixing, nutrient transport, and the rate of photosynthesis at intermediate Reynolds numbers \cite{Santhanakrishnan2012, Genin2013,Miller2017b}. A better understanding of the velocity fluctuations in suspensions may also be of use in the design of biomimetic systems for related purposes \cite{Priya2011,Priya2012,Parker2012},

One key to developing analytical estimates for velocity fluctuations is to start with a tractable `building block,' in this case a simple model for a vortex ring.  There exists a great wealth of literature containing analytical, numerical, and experimental results for vortex rings \cite{maxworthy_structure_1972,cantwell_decay_1988,stanaway_numerical_1988,shariff_vortex_1992,saffman,Cater2004,fukumoto_global_2008,fukumoto_global_2010,dabiri_fluid_2004,dabiri_note_2006,shadden_lagrangian_2006,delbende_dynamics_2009}, but to study the role of viscous vortex decay, a classical ideal vortex model is insufficient.  Instead, we shall use an intermediate-Reynolds number model of a decaying vortex ring due to \citet{fukumoto_global_2008}.

In the following pages we show analytically and verify numerically that the probability distribution for the velocity fluctuations of a dilute suspension of vortex rings is a truncated $5/3$-stable distribution which decays like $\lvert\uc\rvert^{-8/3}$ for a component of velocity $\uc$. These results are robust in the sense that any flow produced by impulses sufficiently localized in both space and time will produce the same velocity distribution.  The variance of~$\uc$ (mean energy) is shown to be linear in the vortex volume fraction $\volfrac$ as expected from such a superposition of individual velocity fields.  However, the width of the core scales as~$\volfrac^{3/5}$ rather than $\volfrac^{1/2}$, suggesting that the tails of the distribution contribute at leading-order to the energy.

The paper is structured as follows. In Section \ref{SingleVR}, we present a model of a viscous vortex ring due to \citet{fukumoto_global_2008} and analyze the moments of the resulting flow field. In Section \ref{Energy}, we build a suspension of viscous vortices by superimposing the flow fields of individual model vortex rings, and we subsequently derive an estimate for the energy of the suspension. This analysis is expanded in Section \ref{Distribution} to determine the full velocity distribution analytically. These findings are confirmed numerically using simulations involving the evaluation of transient velocity fields over multiple scales. We show in Section \ref{Robustness} that under a particular set of conditions, the $\lvert\uc\rvert^{-8/3}$ power law observed in the distribution is robust and is a consequence of swimming occurring in a three-dimensional fluid  Concluding remarks are given in Section \ref{discussion}.

\section{A single viscous vortex ring}
\label{SingleVR}

\subsection{Model}
\label{ModelVR}

\begin{figure}
	\centering
	\includegraphics[width=.32\textwidth]{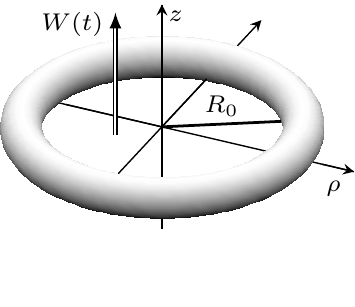}
	\hspace{.002\textwidth}
	\includegraphics[width=.32\textwidth]{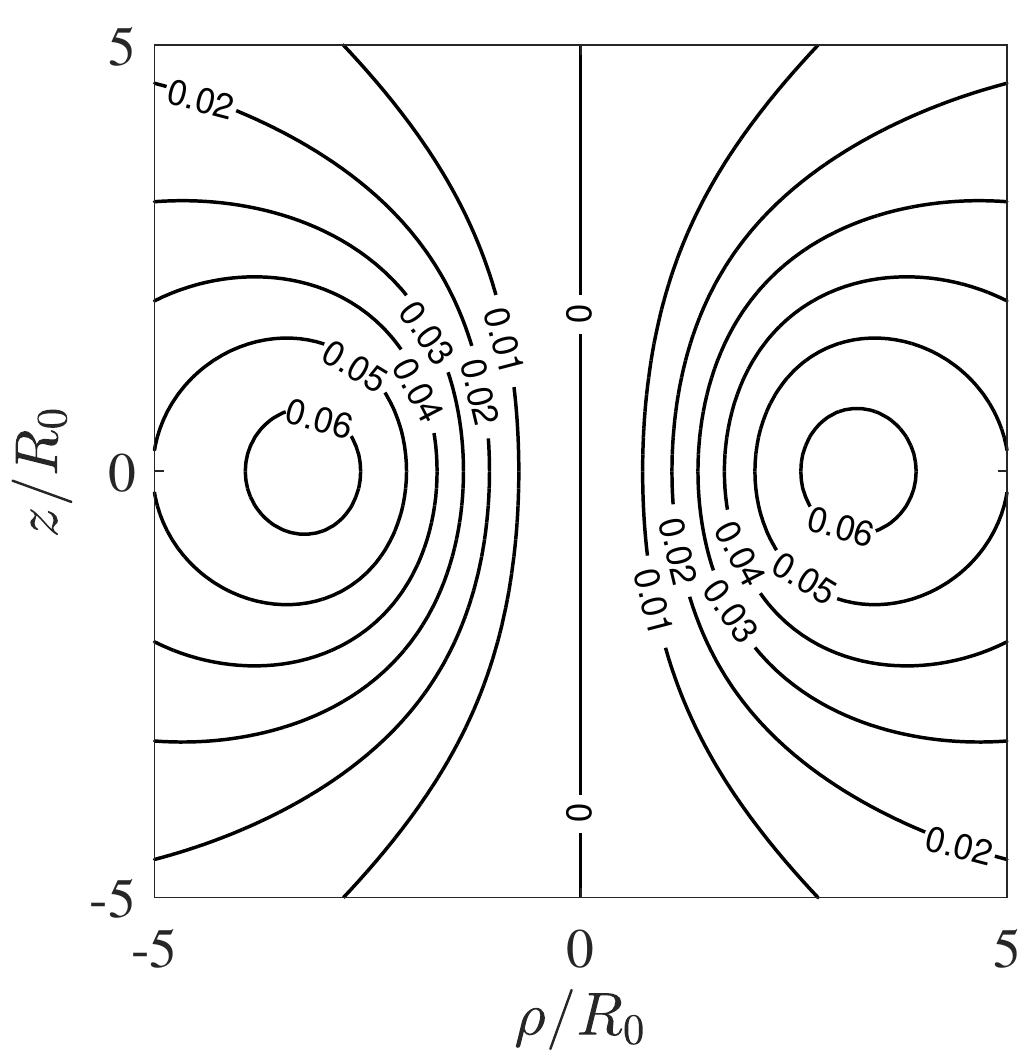}
	\hspace{.002\textwidth}
	\includegraphics[width=.32\textwidth]{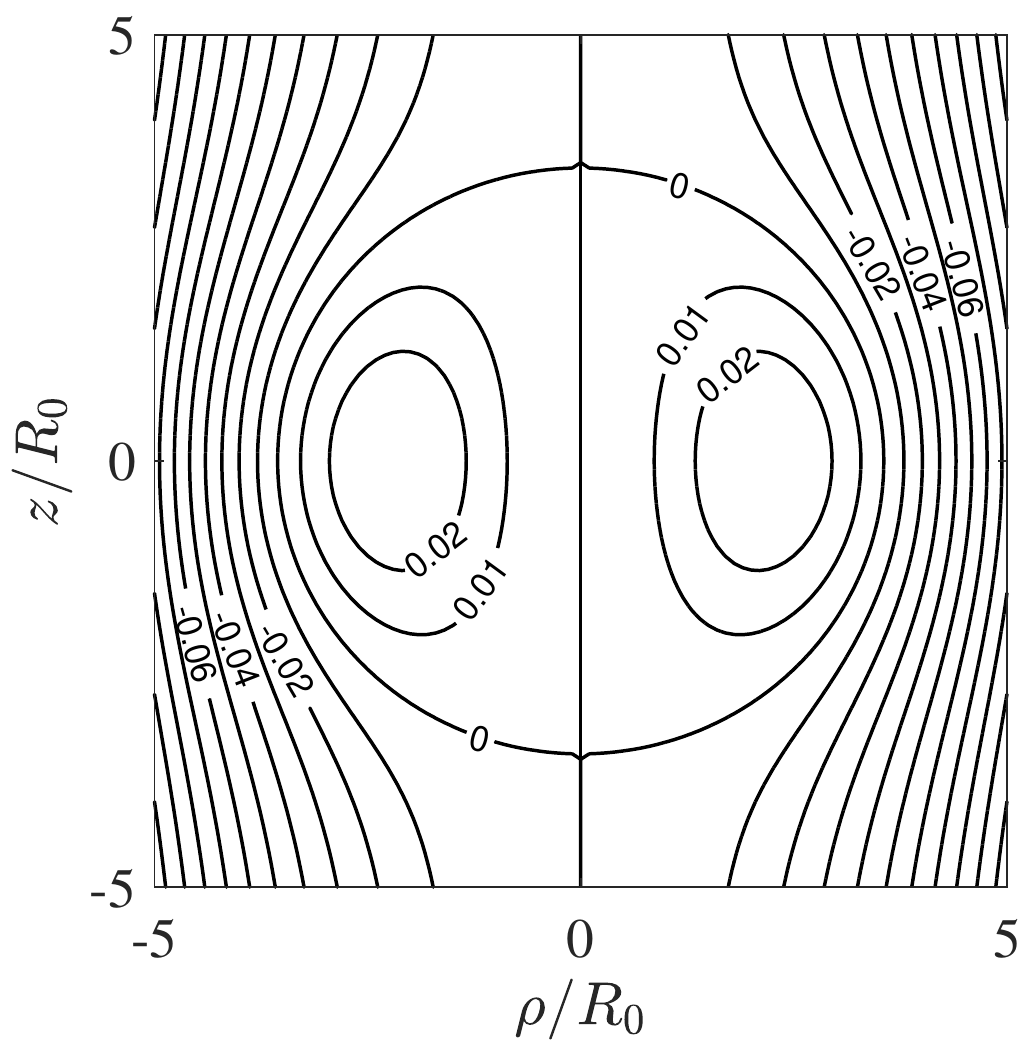}
	\caption{(Left) Diagram of an early-stage vortex ring.  (Center) Contours of the streamfunction $\Psi$ normalized by $\Gamma_0 R_0$ in the lab frame from Eq.~\eqref{eq:IC_streamfunction} at time $\time = R_0^2/\nu$.  (Right) The same normalized streamfunction in a frame moving with the vortex ring.}
	\label{fig:streamfunction}
\end{figure}

\begin{figure}
	\centering
	\includegraphics[width=.49\textwidth]{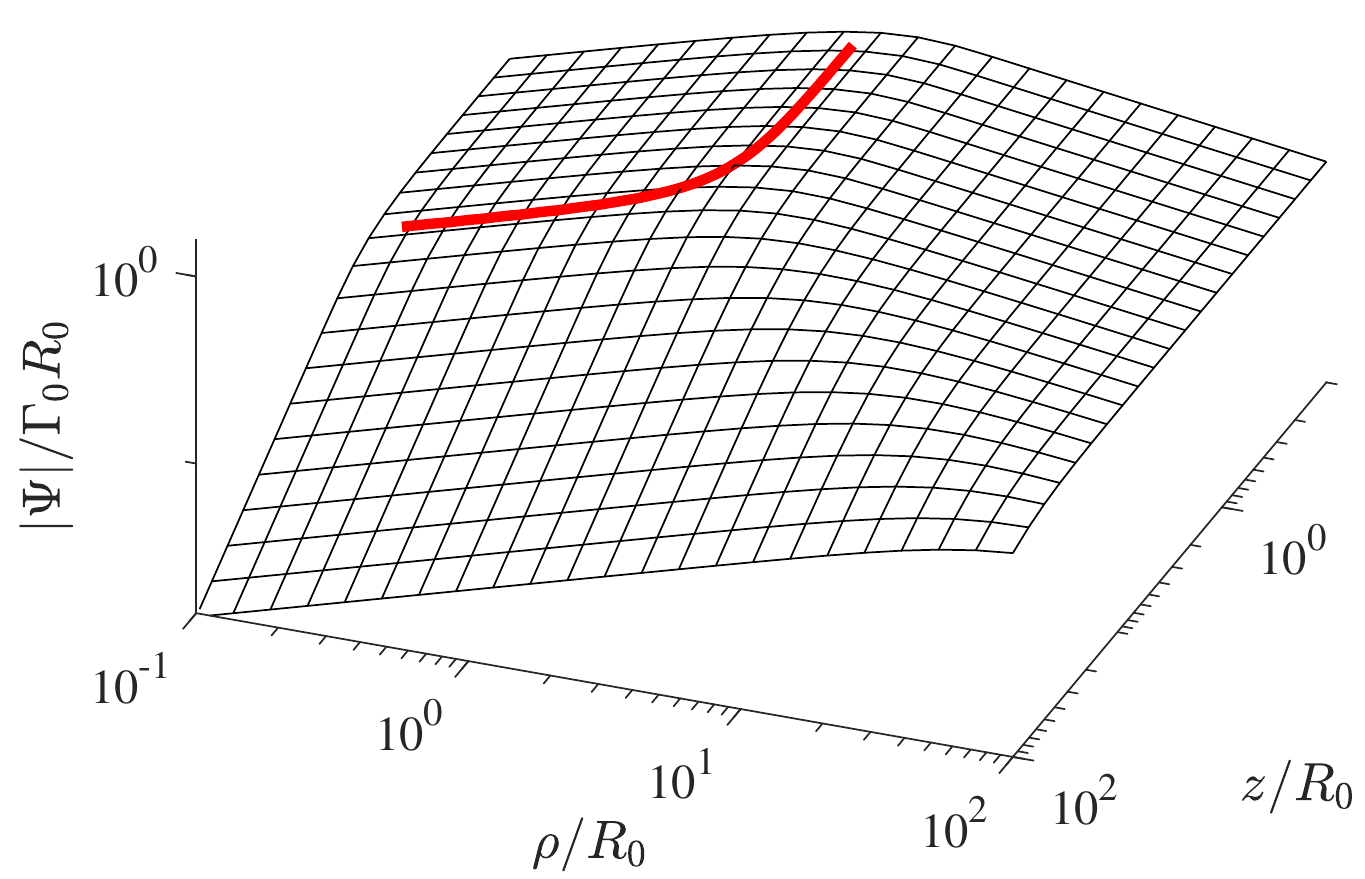}
	\hspace{.002\textwidth}
	\includegraphics[width=.49\textwidth]{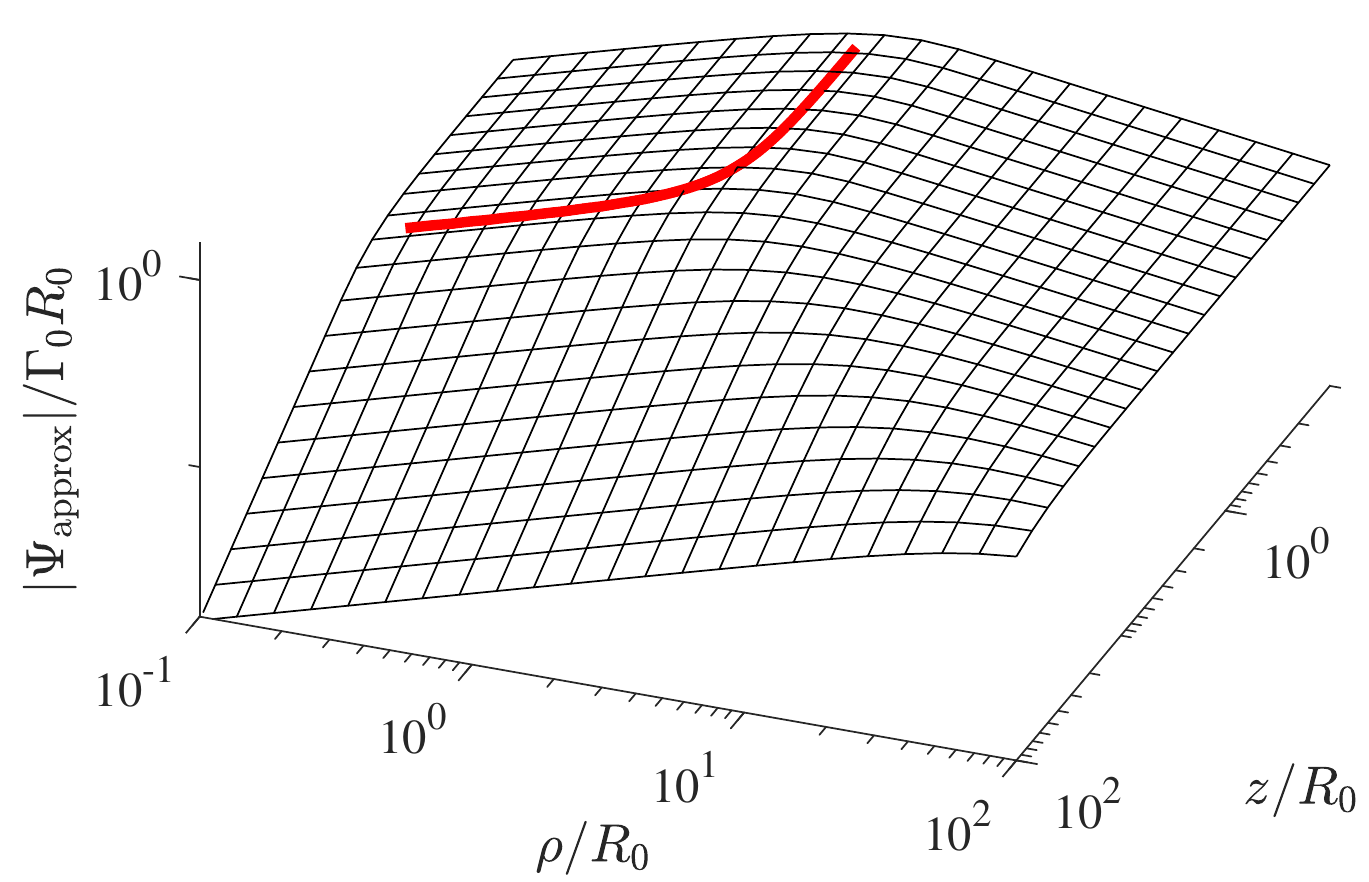}
	\hspace{.002\textwidth}
	\includegraphics[width=.49\textwidth]{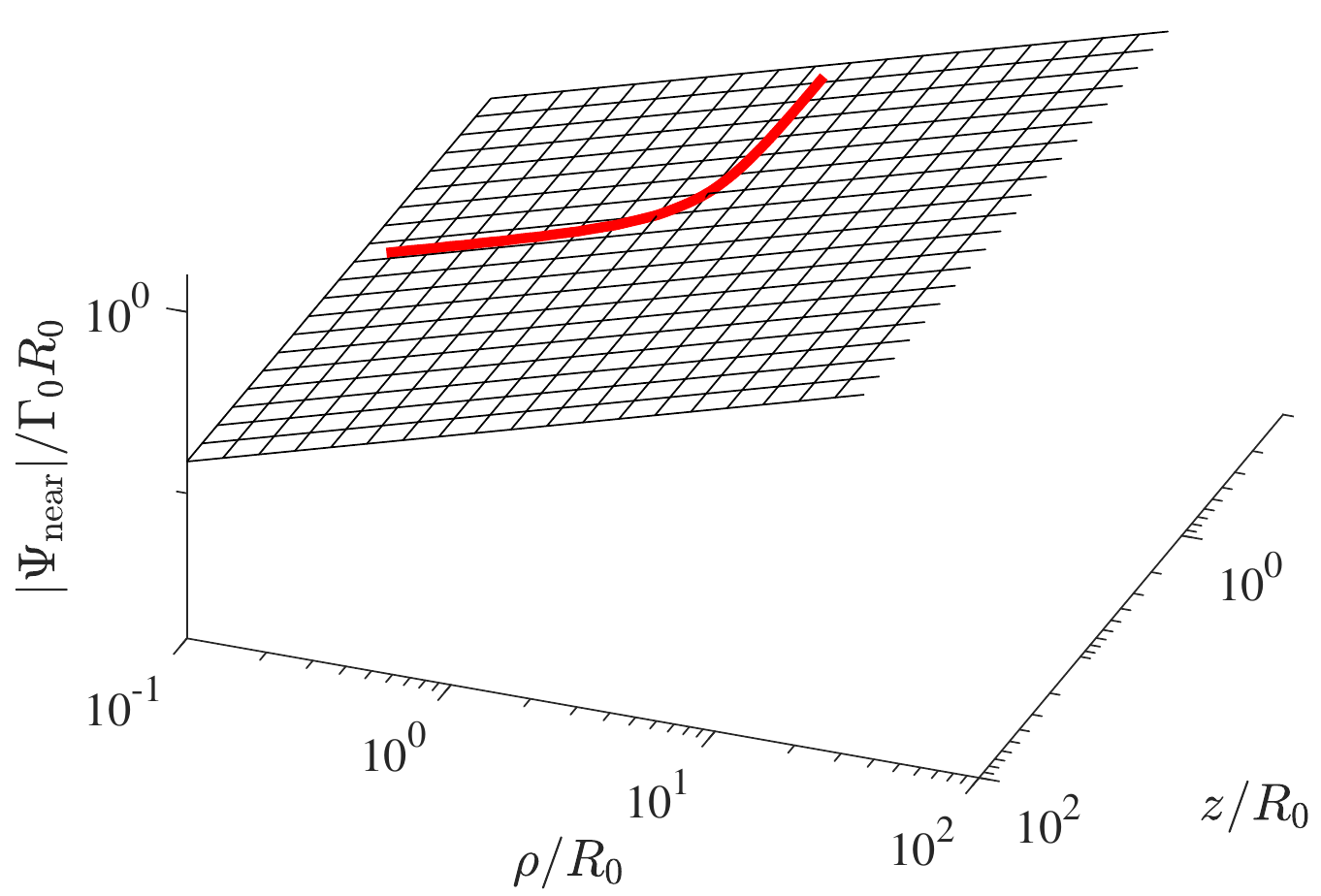}
	\hspace{.002\textwidth}
	\includegraphics[width=.49\textwidth]{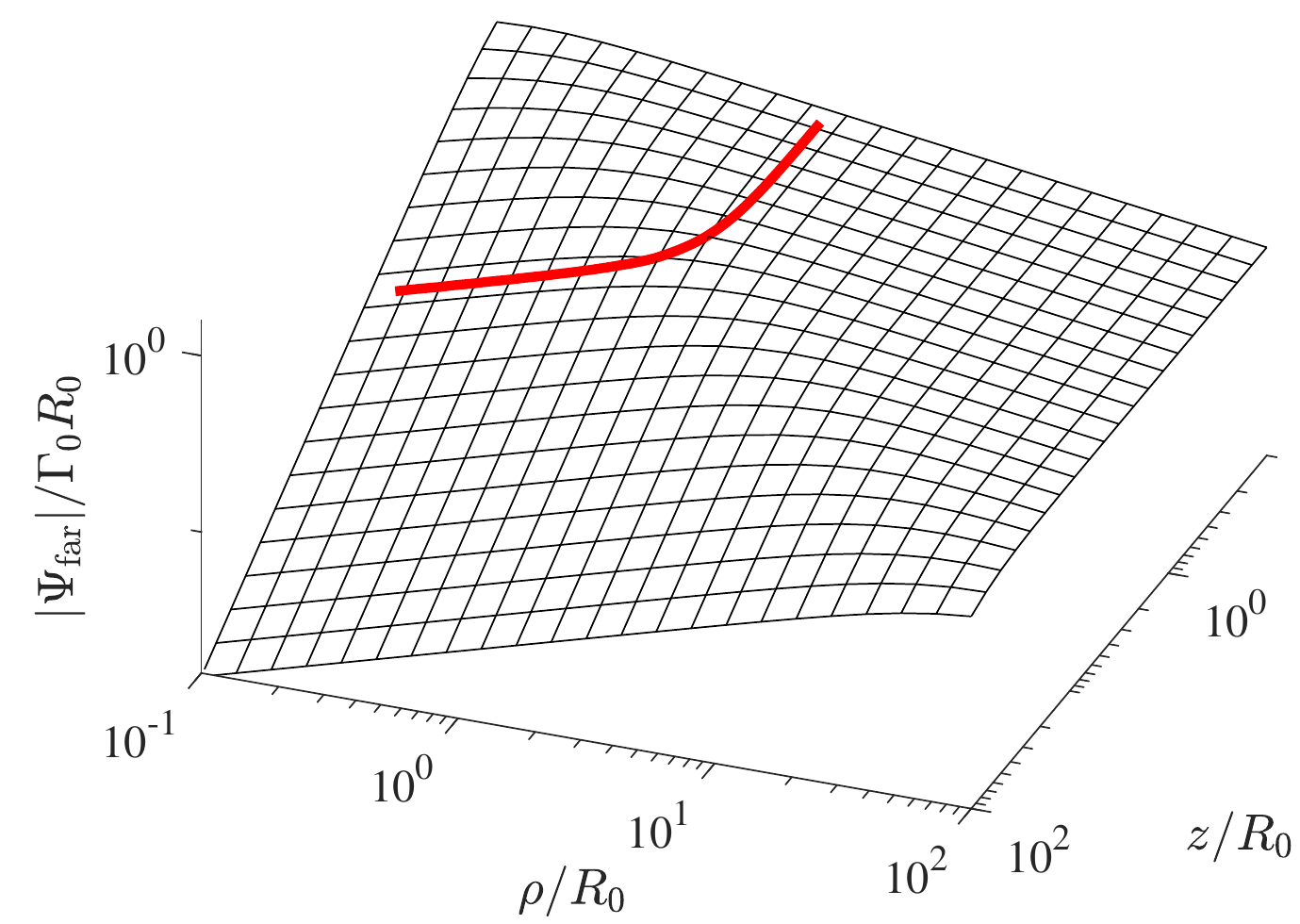}
	\caption{With $\nu\time/R_0^2 = 1$, plots of $|\Psi|$, its approximation $|\Psi_\textup{approx}|$ from \eqref{psiApproximation}, and the near-field and far-field approximations corresponding to the velocity fields from \eqref{BasicVelApprox}, all normalized by $\Gamma_0 R_0$. We see excellent agreement between $\Psi$ and $\Psi_{\textup{approx}}$ uniformly in space, while $\Psi_\textup{near}$ is very similar inside the thick, solid curve ($\xi = 1$) and $\Psi_\textup{far}$ is very similar outside the bold curve.}
	\label{fig:psiPlots}
\end{figure}

Before analyzing a suspension of vortices, we start by presenting a model of a single viscous vortex ring due to \citet{fukumoto_global_2008}. They consider the case of an axisymmetric vortex filament with initial azimuthal vorticity
\begin{equation}
\label{initial_vorticity}
\zeta(\rho,z,\time=0)
=
\Gamma_0\, \delta(z)\, \delta(\rho-R_0),
\end{equation}
where $\delta$ is the Dirac delta function, $\Gamma_0$ is the initial circulation, $R_0$ is the initial radius of the vortex ring, $\rho$ and $z$ are the radial and axial directions in space relative to the vortex ring (see the diagram in Figure \ref{fig:streamfunction}), and $\time$ is time.
In this setting it is convenient to define a streamfunction $\Psi(\rho,z,\time)$, where the velocity in the lab frame is given by $\vv = \rho^{-1}\, \nabla^\perp \Psi$, with $\nabla^\perp = \hat{z}\,\partial_\rho - \hat{\rho}\,\partial_z$.
Defining the Reynolds number as $\Rey := \Gamma_0/\nu$, where $\nu$ is the kinematic viscosity, \citet{fukumoto_global_2008} find that the swirl-free flow, to leading order in small Reynolds number with initial condition (\ref{initial_vorticity}), takes the form:
\begin{subequations}
\begin{align}
  \label{eq:IC_vorticity}
  \zeta(\rho,z,\time) &= \displaystyle \frac{\Gamma_0 R_0}{4\sqrt{\pi} (\nu\time)^{3/2}} \exp\!\left( - \frac{z^2 + \rho^2 + R_0^2}{4\nu\time} \right) I_1\!\left( \frac{R_0 \rho}{2\nu\time} \right), \\
  \Psi(\rho,z,\time)
  &=
  \tfrac14\Gamma_0 R_0 \rho \int_0^\infty
  \left[
    \ee^{mz}\,\erfc\!\left( \frac{2m\nu\time+z}{2\sqrt{\nu\time}} \right)
    +
    \ee^{-mz} \erfc\!\left( \frac{2m\nu\time-z}{2\sqrt{\nu\time}} \right)
  \right]
  J_1\!\left( mR_0 \right) J_1\!\left( m\rho \right) \!\dint m.
  \label{eq:IC_streamfunction}
\end{align}
\label{eq:IC}
\end{subequations}
Here~$J_1$ and~$I_1$ are standard and modified Bessel functions of the first kind, respectively, and $\erfc$ is the
complementary error function. The circulation is found to decay in time as $\Gamma(\time) = \displaystyle \Gamma_0 \left[ 1 - \exp\!\left( -{R_0^2}/{4\nu\time} \right) \right]$.
A useful approximation to $\Psi$ is
\begin{equation}
	\Psi(\rho,z,\time)
	\approx
	\frac{\Gamma_0 R_0^2}{2\sqrt{\pi}}
	\left( \int_0^{\xi} \ee^{-{\xi^\prime}^2} d\xi^\prime - \xi \ee^{-\xi^2} \right)
	\frac{\rho^2}{(z^2+\rho^2)^{3/2}}, \qquad
	\frac{R_0}{\sqrt{4\nu\time}} \ll \max(\xi,1)\,,
	\label{psiApproximation}
\end{equation}
where
\begin{equation}
  \xi(\rho,z,\time) \ldef \sqrt{\nofrac{(z^2+\rho^2)}{4\nu \time}}
  \label{eq:xi}
\end{equation}
is a dimensionless measure of the position relative to the `viscous front' at~$\xi=1$ associated with the outward propagation of viscous
stresses.  Crucially, the
form~\eqref{psiApproximation} is valid even at small~$\time$, as long as we
are considering points well outside the vortex ring.  Applying small and large $\xi$ approximations to~\eqref{psiApproximation}, we find an approximate velocity field
\begin{equation}
	\label{BasicVelApprox}
	\vv(\rho,z,\time) = \begin{cases}
		\cfrac{\Gamma_0 R_0^2 \,\hat{{\bm z}}}{12\sqrt{\pi} (\nu \time)^{3/2}} & \xi \lesssim 1, \\[12pt]
		\cfrac{\Gamma_0 R_0^2 [(2z^2-\rho^2) \,\hat{{\bm z}} + 3z\rho\, \hat{{\bm \rho}}]}{4 (z^2+\rho^2)^{5/2}} & \xi \gtrsim 1,
	\end{cases}
        \qquad
	\frac{R_0}{\sqrt{4\nu\time}} \ll \max(\xi,1)\,,
\end{equation}
with a relatively sharp transition region around the viscous front $\xi = 1$. Note that although the two parts of \eqref{BasicVelApprox} were derived in the asymptotic regimes where $\xi \ll 1$ and $\xi \gg 1$, respectively, they are good approximations for nearly all points (see Figure \ref{fig:psiPlots} for a comparison between the full stream-function and the near- and far-field approximations), except for small times and right at the viscous front $\xi = 1$.

The vortex ring also propels itself forward in time. To find the self-advection of the vortex ring and incorporate it into the model, Fukumoto and Kaplanski use the Helmholtz--Lamb transformation, from which they determine the instantaneous speed $\Wc(\time)$ of the vortex ring and the net displacement $S(\time)$ in the positive $z$-direction \citep{fukumoto_global_2008}. Incorporating the vortex speed $\Wc$ into the streamfunction by subtracting $\frac12\rho \Wc^2$ from $\Psi$ results in the more familiar ellipsoidal envelope corresponding to $\Psi = 0$ as shown in Figure \ref{fig:streamfunction} (right).

The model matches previous estimates for the early and late time velocities \citep{cantwell_decay_1988,saffman_vortex_1970,saffman}. \citet{fukumoto_global_2008} also validate their model against experimental results from \citet{Cater2004} with $\Rey = 2000$ and find excellent agreement, suggesting \eqref{eq:IC} accurately captures the structure of the fluid flow for a broad range of intermediate Reynolds numbers, including those of various jellyfish \cite{McHenry2003,Dabiri2014,Dabiri2015}. For the {\it Aurelia aurita} jellyfish in a Danish fjord studied by \citet{Olesen1994}, we can estimate that $\Rey$ ranges from around $60$ to $2160$.

\subsection{Moments of the velocity distribution}
\label{Moments}

In this section, we study the moments of the velocity field associated with a single vortex ring integrated over both time and space:
\begin{equation}
\label{eq:momentsDefinition}
M_n := \int_0^\infty \int_V | \vv |^n\,\dint V \dint \time,
\end{equation}
where $V$ is our domain, in this section taken to be $\mathbb{R}^3$. At the outset, it is not clear which moments exist, if any, and we shall see that many do not.
To this end, we use \eqref{BasicVelApprox} to approximate the far-field velocity field and see that $| \vv |$ decays like $r^{-3}$ (where $r = \sqrt{z^2 + \rho^2}$) as $r \rightarrow \infty$. Upon integrating over space, we therefore have that
\begin{equation}
\label{eq:far-field_moment_integral}
\int_V | \vv |^n\,dV = \int_V | \vv |^n\,r^2 \,\dint r \,\dint\Omega
\end{equation}
is infinite for $2-3n \ge -1$, which means that $M_n = \infty$ for all $n \le 1$,
and for $n > 1$,
\begin{equation}
\label{eq:far-field_moment_integral_time_series}
\int_V | \vv |^n~dV = O(\Gamma_0^n (\nu t)^{3(1-n)/2}),
\end{equation}
valid as $t \rightarrow \infty$.

Another possible source of moment divergence lies at time $t = 0$, when the velocity field is singular at the vortex core. For small times the evolution of vorticity near a point on the vortex ring may be studied using a line vortex approximation. Consider then a line vortex located at the origin; the vorticity $\zeta$ is the Green's function for the heat equation, multiplied by the initial circulation:
\begin{equation}
\label{eq:viscous_line_vortex_vorticity}
\zeta = \frac{\Gamma_0}{4\pi \nu \time} \exp \left( -\frac{x^2+y^2}{4\nu \time} \right).
\end{equation}
Then the swirl velocity is
\begin{equation}
\label{eq:viscous_line_vortex_velocity}
\vc
=
\frac{\Gamma_0}{2\pi \sqrt{x^2+y^2}}
\left[ 1 - \exp \left( -\frac{x^2+y^2}{4\nu \time} \right) \right],
\end{equation}
counterclockwise around the origin. Near the vortex, $v \approx \frac{\Gamma_0 \sqrt{x^2+y^2}}{8 \pi \nu t}$. Integrating over a finite neighborhood around the origin, we see that
\begin{equation}
\label{eq:near_space_integral_moments}
\int | \vv |^n \,\dint A = O(\Gamma_0^n (\nu \time)^{1-n/2}),
\end{equation}
valid as $t \downarrow 0$, is finite for all nonnegative $n$ and positive $\time$, so this region does not contribute to any possible divergence of $\int_V | \vv |^n\,\dint V$ for any $n \ge 0$.

Looking across the entirety of the spatial domain, the arguments above suggest the existence of $\int_V | \vv |^n\,\dint V$ for all $n > 1$, but we are particularly interested in the moments $M_n$, which are integrals over both space and time. Examining the rate of decay of \eqref{eq:far-field_moment_integral_time_series} for large times results in infinite moment $M_n$ precisely when $3(1-n)/2 \ge -1$, or $n \le \frac{5}{3}$. Similarly, behavior of \eqref{eq:near_space_integral_moments} at small times results in infinite moment $M_n$ when $1-n/2 \le -1$, or $n \ge 4$. Thus, {\it moments of $\vv$ exist only for $\tfrac{5}{3} < n < 4$.}

\begin{figure}
	\centering
	\includegraphics[width=.49\textwidth]{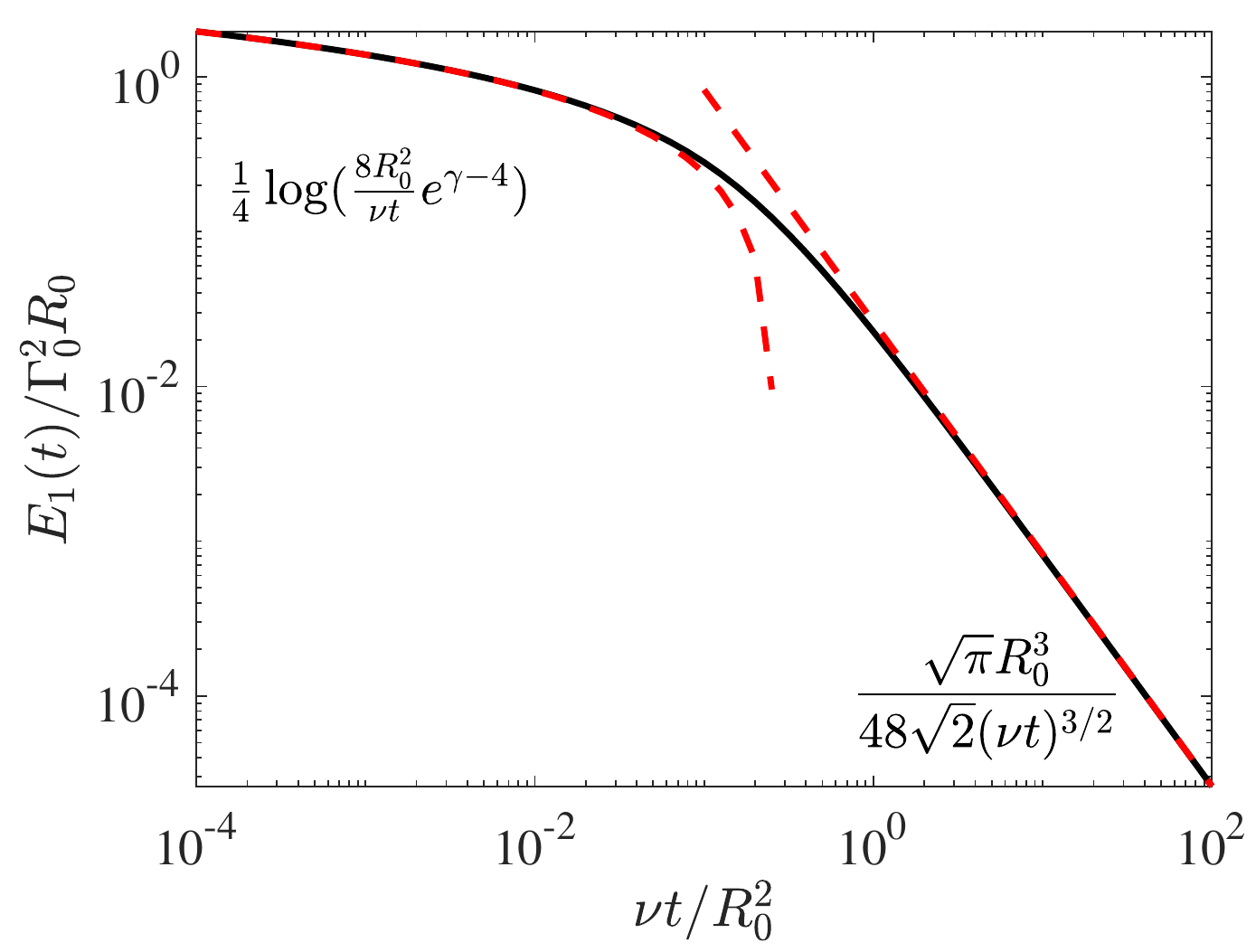}
	\caption{The energy integrated over all space $E_1(\time)$ for a single vortex ring normalized by $\Gamma_0^2 R_0$ and compared with the small and large-time asymptotics in \eqref{eq:energyAsymptotics}.}
	\label{fig:energy_asymptotics}
\end{figure}

In particular, the variance $M_2$ is finite, which has important consequences both mathematically and physically. The energy in the entire fluid at a time $\time$, in the Fukumoto and Kaplanski model, is given by \citep{fukumoto_global_2008}
\begin{equation}
\label{eq:energy}
E_1(\time)
=
\tfrac12 \int_V | \vv |^2~\dint V
=
\frac{\sqrt{\pi}\, \Gamma_0^2 R_0^4}{48 \sqrt{2} (\nu\time)^{3/2}}
\,{}_2F_2\!\left( \tfrac32, \tfrac32; \tfrac52, 3; -\tfrac{R_0^2}{2\nu\time} \right),
\end{equation}
where $_2F_2$ is a generalized hypergeometric function.
This has asymptotic forms
\begin{equation}
\label{eq:energyAsymptotics}
E_1(\time) \sim \begin{cases}
\Gamma_0^2 R_0 \left( \tfrac14\log(8 R_0^2 / \nu\time) + \tfrac14\gamma -1 \right), \quad & \textup{as } \time \downarrow 0, \\
\frac{\sqrt{\pi}\, \Gamma_0^2 R_0^4}{48 \sqrt{2} (\nu\time)^{3/2}}\,, & \textup{as } \time \rightarrow \infty,
\end{cases}
\end{equation}
where $\gamma$ is the Euler--Mascheroni constant. These asymptotic forms are plotted in Figure \ref{fig:energy_asymptotics} to indicate their degree of accuracy when compared to $E_1$.  Remarkably, $E_1(\time)$ can be integrated over time exactly, to obtain the total vortex energy
\begin{equation}
\label{eq:SingleVREnergyContribution}
\mathcal{E}_1
:= \tfrac12 M_2 =
\int_0^\infty
E_1(\time)
\,\dint\time
=
\frac{\Gamma_0^2 R_0^3}{6 \nu}
=
\tfrac16\Rey^2\,\nu R_0^3.
\end{equation}

\section{Energy of a suspension of viscous vortices}
\label{Energy}

In this section we find an analytical estimate for the energy of a suspension of viscous vortices, which will be used in the analysis of the full velocity distribution.
Vortex rings are assumed to come into being uniformly in time, space, and orientation, into an otherwise quiescent infinite bath.  The rate of vortex production is $\prodrate$ vortices per unit time per unit volume, or with dimensionless birth rate $\volfrac := \prodrate R_0^5 / \nu$. In nature, concentrations {\it Aurelia aurita} jellyfish have been observed in the range of $1 \times 10^{-6}$ to $3 \times 10^{-4}$ medusae per cubic centimeter with mean radius $R_0$ ranging from $0.125$\,cm to $2.7$\,cm depending on the time of year \citep{Olesen1994}. Meanwhile, \citet{McHenry2003} found that jellyfish pulsed at a rate of once per second for smaller medusae, and once per two seconds for larger medusae. We therefore estimate that, for the suspension of vortices, $\volfrac$ ranges from $3 \times 10^{-8}$ in early spring to $0.3$ in late summer. Thus, we will assume that $\volfrac \ll 1$, and therefore that any vortex-vortex interactions are negligible.

Consider the velocity field~$\vv(\rv,\time) = \rho^{-1}\, \grad^\perp \Psi$ for a vortex initially at the origin and pointing in the~$\zuv$ direction, as in Figure \ref{fig:streamfunction}. Rotating and translating the velocity to represent a vortex with arbitrary position and direction, we first obtain the rotated velocity field
\begin{equation}
\Rot\cdot\vv\!\l(\Rot^{-1}\cdot\rv,\time\r)
\!,
\end{equation}
where $\Rot$ is a rotation matrix, and then translate the field to point~$\Rv$ (replacing~$\rv$ by~$\rv-\Rv$):
\begin{equation}
\Rot\cdot\vv\!\l(\Rot^{-1}\cdot(\rv - \Rv),\time\r)
\!.
\end{equation}
Writing the vortex position in time as
\begin{equation}
\Rv(\time)
=
\Rv(0) + \Ss(\time)\,\Rot\cdot\zuv,
\qquad
\Ss(0) =0,
\end{equation}
(recall that $S(\time)$ is the vortex displacement and $W(\time) = S^\prime(\time)$ is the speed),
thus results in its induced velocity field
\begin{equation}
\Rot\cdot\vv\!\l(\Rot^{-1}\cdot(\rv - \Rv(0)) - \zuv\Ss(\time)\,,\,\time\r).
\end{equation}

Summing the velocity contributions at a point $\rv_0$ from $\numV$ independent vortices, which are initially located at random points $\Rv_k$, results in 
\begin{equation}
\U
=
\sum_{k=1}^{\numV}
\Rot_k\cdot
\vv\!\l(\Rot_k^{-1}\cdot(\rv_0 - \Rv_k) - \zuv\,\Ss(\T_k)\,,\T_k\r),
\label{eq:V0many}
\end{equation}
where the random variable $\T_k$ denotes the age of the $k$th vortex, and $\Rot_k$ is a random rotation matrix, which enforces isotropy.  We assume~$\numV = \prodrate V\tau$ is constant, where~$V$ is the total volume of the domain and~$\tau$ is the lifetime of a vortex. Here, $V,\tau$ are assumed finite, but we will examine the infinite volume and time limits shortly.

The expected value of~$\U$, $\langle\U\rangle$, averaged over all positions, orientations, and birth times, is
\begin{equation}
\langle\U\rangle
=
\numV
\int_\Omega
\int_0^\tau
\int_V
\Rot\cdot\vv\!\l(\Rot^{-1}(\Omega)\cdot(\rv_0 - \rv) - \zuv\,\Ss(\time)
\,,\,\time\r)
\,\frac{\!\dint V_{\rv}}{V}
\,\frac{\!\dint\time}{\tau}
\,\frac{\!\dint\Omega}{4\pi},
\end{equation}
with~$\Omega$ the solid angle that determines the rotation matrix. With the change of variables
\begin{equation}
\rv' = \Rot^{-1}(\Omega)\cdot(\rv_0 - \rv) - \zuv\,\Ss(\time),
\qquad \time'=\time,
\label{eq:rvprime}
\end{equation}
we have~$\pd\rv'/\pd\rv = -\Rot^{-1}(\Omega)$,
and~$\pd\rv'/\pd\time = -\Wc(\time)\zuv$.  The Jacobian matrix for the transformation
is
\begin{equation}
\frac{\pd(\rv',\time')}{\pd(\rv,\time)}
=
\begin{pmatrix}
-\Rot^{-1}(\Omega) & -\Wc(\time)\zuv \\ 0 & 1
\end{pmatrix}
\end{equation}
with determinant~$-1$, so the Jacobian does not modify the integral:
\begin{equation}
\langle\U\rangle
=
\numV
\int_\Omega
\int_0^\tau
\int_{V'(\rv_0,\time',\Omega)}
\Rot\cdot\vv(\rv',\time')
\,\frac{\!\dint V_{\rv'}}{V}
\,\frac{\!\dint\time'}{\tau}
\,\frac{\!\dint\Omega}{4\pi}.
\end{equation}
Here~$V'(\rv_0,\time',\Omega)$ is the domain of integration transformed according to~\eqref{eq:rvprime}.

Similarly, the~$q$th absolute moment of~$\U$ can be computed as
\begin{equation}
\langle|\U|^q\rangle
=
\numV
\int_\Omega
\int_0^\tau
\int_V
|\vv(\rv',\time')|^q
\,\frac{\!\dint V_{\rv'}}{V}
\,\frac{\!\dint\time'}{\tau}
\,\frac{\!\dint\Omega}{4\pi}.
\end{equation}
Integrating over the orientation angles and dropping the primes,
\begin{align}
\label{eq:Evq}
\langle|\U|^q\rangle
&=
\numV
\int_0^\tau
\int_V
|\vv(\rv,\time)|^q
\,\frac{\!\dint V_{\rv}}{V}
\,\frac{\!\dint\time}{\tau} \\
&=
\prodrate
\int_0^\tau
\int_V
|\vv(\rv,\time)|^q
\dint V_{\rv}\,
\dint\time.
\nonumber
\end{align}

Setting $q = 2$, taking $V = \mathbb{R}^3$ and $\tau \rightarrow \infty$ (and dividing by two), we find the expectation of the energy
\begin{equation}
\langle E \rangle
=
\tfrac12\prodrate
\int_0^\tau
\int_V
|\vv(\rv,\time)|^2
\dint V_{\rv}
\dint\time
=
\prodrate \mathcal{E}_1 = \frac\volfrac6\frac{\Gamma_0^2}{R_0^2}.
\label{eq:EE}
\end{equation}
Thus, the expected energy is $\mu$ times the energy of a single vortex integrated over time and space.  This is reasonable: in this noninteracting dilute limit, the energy of the system is the sum of the energy of the individual vortices.

\section{Velocity distribution}
\label{Distribution}

A more refined analysis than that of Section~\ref{Energy} allows us to characterize the entire velocity distribution, rather than just the moments.  This clarifies whether the dominant contribution to the moments arises from near or far field dynamics, as well as facilitating potential comparisons to experiments.  For small concentrations, we will find stable distributions similar to \citet{Zaid2011} for suspensions of microswimmers, though the relationship between spatial velocity decay and the tail exponents is modified here by the additional temporal behavior of the vortices.

\subsection{Single vortex}
\label{Single_VR_pdf}

We first consider the velocity distribution due to a single vortex ring, which
will be used in Section \ref{Suspension_pdf} to derive the marginal
distribution for the velocity fluctuations in a suspension
of viscous vortices. We choose a random point~$\rv = \rv_0 + (\rho \cos \theta, \rho \sin \theta, z)$ uniformly inside the ball~$V = B_L(\rv_0)$ of radius $L$ centered at $\rv_0$, and choose a random vortex age $\time$ uniformly in $[0,\tau]$.   The probability density function $\prob_{U^1}(\uc)$ for the magnitude of the single-vortex velocity $U^1 = \lvert\U^1\rvert$ is
\begin{equation}
\label{pdf_velocity_single_vr}
\prob_{U^1}(\uc)
=
\int_0^\tau \int_V \delta(\uc-\vc(\rv,\time)) \frac{\dint V_{\rv}}{V} \frac{\dint t}{\tau}
\end{equation}
where~$\vc(\rv,\time) = | \vv(\rv,\time)|$. The delta function constrains the integral to a hypersurface~$\vc(\rv,\time) = \uc$:
\begin{equation}
\label{pdf_velocity_single_vr_surface}
\prob_{U^1}(\uc)
=
\frac{1}{V\tau} \int_{\vc(\rv,\time) = \uc}
\frac{1}{| \grad_{(\rv,\time)} \vc(\rv,\time) |}
\,\dint S_{\rv,\time}
\end{equation}
where~$| \grad_{(\rv,\time)} \vc(\rv,\time) |$ is a Jacobian~\cite{Hormander}.
An analytical estimate may be achieved by splitting the integral into two pieces, $\xi \le 1$ and $\xi \ge 1$ with~$\xi = |\rv|/\sqrt{4\nu\time}$, and
using \eqref{BasicVelApprox}, valid for small $\uc$, to approximate the velocity.  (We neglect the transition region near~$\xi=1$.)  This straightforward but somewhat messy calculation is carried out in Appendix~\ref{Single_VR_pdf_Computation}.  By combining Eqs.~\eqref{pdf_velocity_late_time_single_vr} and~\eqref{pdf_velocity_xi_ge_1_single_vr}, we find that
\begin{equation}
\label{ppdf}
\prob_{U^1}(\uc)
\lesssim
\frac{0.1959}{V\tau} \frac{\Gamma_0^{5/3} R_0^{10/3}}{\nu}\,\uc^{-8/3},
\qquad
\ucmin \ll \frac{\uc R_0}{\Gamma_0} \ll 1,
\end{equation}
where
\begin{equation}
  \ucmin
  =
  \frac{R_0^3}{\min(V,(\nu\tau)^{3/2})}.
\end{equation}
The approximation breaks down as~$(\uc R_0/\Gamma_0) \uparrow 1$ because then the details of the near field of the vortex become important, and we cannot use~\eqref{BasicVelApprox} to go from~\eqref{pdf_velocity_single_vr_surface} to~\eqref{ppdf} as we did above.  The approximation~\eqref{ppdf} also breaks down as~$(\uc R_0/\Gamma_0) \downarrow \ucmin$ because, at fixed~$V$ and~$\tau$, the region~$\vc(\rv,\time) < \ucmin\Gamma_0/R_0$ falls outside the domain of integration in~\eqref{pdf_velocity_single_vr}.  The value of~$\ucmin$ is typically small, indicating a wide range of validity for~\eqref{ppdf}, as long as the domain radius~$L$ is much larger than the vortex size~$R_0$, and the time of integration~$\tau$ is much longer than the viscous dissipation time~$R_0^2/\nu$.

\begin{figure}
	\centering
	\includegraphics[width=0.49\textwidth]{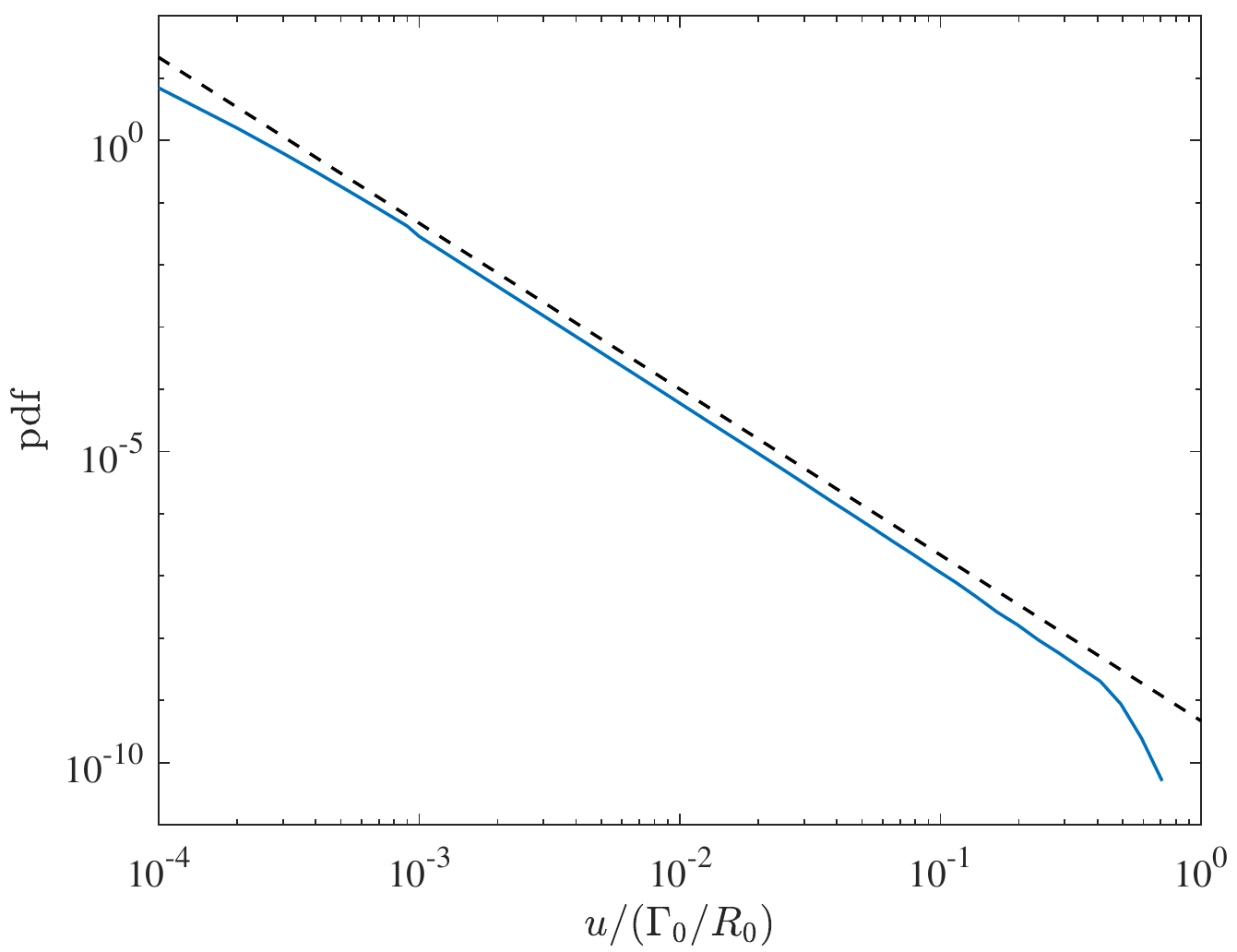}
	\caption{The numerically evaluated velocity probability density function for a single vortex ring (solid line) compared with the analytic approximation \eqref{ppdf} (dashed line). The approximation is about 40\% higher than the numerical values on the segment with $0.001 \Gamma_0 / R_0 \apprle \uc \apprle 0.04 \Gamma_0 / R_0$.}
	\label{fig:singleVR_PDF}
\end{figure}

In order to probe the accuracy of this approximation, we computed \eqref{pdf_velocity_single_vr} via Monte Carlo integration, finding the velocity at a point $\rv_0$ using a second-order finite difference approximation of \eqref{eq:IC_streamfunction} for a single vortex ring (see Section \ref{numeric_results} for more details) positioned randomly in $B_L(\rv_0)$ with $\Gamma_0 = 100 \nu$, $L = 100\, R_0$, and $\tau = 100\, R_0^2 / \nu$, and continuing to sample until the distribution converged.  Figure \ref{fig:singleVR_PDF} shows a comparison between the numerical computation of \eqref{pdf_velocity_single_vr} and the analytical approximation \eqref{ppdf}. We see that the $\uc^{-8/3}$ power law holds over a wide range of values of $\uc$.
The analytical prediction \eqref{ppdf} is about 40\% too large when compared with the numerics due to the transition region around $\xi \approx 1$. However, this error does not affect the exponent in the $-8/3$ power law, just the prefactor.

\subsection{Suspension of vortices}
\label{Suspension_pdf}

We now use the velocity distribution for a single vortex ring to determine the corresponding distribution for a suspension of vortices, modifying the argument of \citet{Thiffeault2015} that characterized the drifts associated with microswimmers. We will use components of the velocity instead of its magnitude, since components can be added together but not magnitudes.  This additivity of velocity is a good approximation at low volume fractions~$\volfrac$. Since we have assumed isotropy of the suspension, there is no loss in generality in considering only a single component of the fluid velocity~$\uv$.

Starting from the single-vortex distribution~$\prob_{U^1}(\uc)$ for the magnitude of velocity, Eq.~\eqref{pdf_velocity_single_vr}, we convert to the distribution for the components with
\begin{equation}
  \prob_{\U^1}(\uv)
  =
  \int_{V}
  \int_{0}^\tau
  \frac{\delta(\uc - v(\rv,t))}{4\pi u^2}
  \,\frac{\dint\time}{\tau}
  \,\frac{\dint V_{\rv}}{V},
\qquad
\uc = |\uv|,
\end{equation}
where we assumed isotropy of~$\uv$.  We then find the marginal distribution for the $\xc$-component of $\uv$, denoted by $\uc_\xc$:
\begin{align}
\prob_{U_x^1}(\uc_\xc)
&=
\int_{-\infty}^\infty\int_{-\infty}^\infty
\prob_{\U^1}(\uv)
\dint u_y\dint u_z \nonumber \\
&=
\int_{V}
\int_{0}^\tau
\int_{-\infty}^\infty\int_{-\infty}^\infty
\frac{\delta(u-v(\rv,t))}{4\pi u^2}
\,\dint u_y
\,\dint u_z
\,\frac{\dint\time}{\tau}
\,\frac{\dint V_{\rv}}{V},
  \label{probDensityVelocityXSingleVR}
\end{align}
where the superscript~$1$ on~$U_x^1$ and~$\U^1$ is a reminder that this is still for a single vortex.
Carrying out the integrals over $u_y$ and $u_z$ yields
\begin{equation}
\label{probDensityVelocityXSingleVRIntOverYandZ}
\prob_{U_x^1}(\uc_\xc)
=
\int_V
\int_0^\tau
\frac{1}{2v(\rv,\time)} \,[v^2(\rv,\time) > \uc_\xc^2]
\,\frac{\dint\time}{\tau}
\,\frac{\dint V_{\rv}}{V},
\end{equation}
where $[A]$ is the indicator function of~$A$, defined as~$1$ if~$A$ is true, and~$0$ otherwise.

In order to determine the distribution for multiple vortex rings, we compute the characteristic function
\begin{equation}
\label{CharFun}
\langle \ee^{\imi kU_x^1} \rangle
=
\int_{-\infty}^{\infty}
\prob_{U_x^1}(\uc_\xc)\, \ee^{\imi k\uc_\xc}
\,\dint \uc_\xc
=
\int_V
\int_0^\tau
\sinc(kv(\rv,\time))
\,\frac{\dint\time}{\tau}
\,\frac{\dint V_{\rv}}{V},
\end{equation}
where $\sinc(x) := \nofrac{\sin x}{x}$ for $x \neq 0$ and $\sinc(0) := 1$. We find that \begin{equation}
\label{CharFunGamma}
\langle \ee^{\imi kU_x^1} \rangle
=
1 - \frac{\gamma(k)}{V\tau},
\end{equation}
where
\begin{equation}
\gamma(k)
:=
\int_V
\int_0^\tau
\left\{ 1-\sinc(kv(\rv,\time)) \right\}\!
\,\dint\time
\,\dint V_{\rv}\,.
\end{equation}

Recall that~$\prodrate$ is the constant rate of production of vortex rings, per unit space and time.  Hence, after a time~$\tau$ we have $\numV = \prodrate V \tau$ independent vortex rings, which together induce a random velocity~$U_x^\numV$ at the origin.  The random variable~$U_x^\numV$ has characteristic function
\begin{equation}
\label{CharFunN}
\langle \ee^{\imi kU_x^\numV} \rangle
=
\langle \ee^{\imi kU_x^1} \rangle^\numV
=
\left( 1 - \frac{\gamma(k)}{V\tau} \right)^{\prodrate V\tau}
\sim
\exp \left( -\prodrate \gamma(k) \right)
\end{equation}
as $V,\tau \rightarrow \infty$ \cite{Thiffeault2015}. Therefore, for the suspension of vortices, the probability density function of velocities is obtained from the inverse Fourier transform
\begin{equation}
\label{probDensityVelocityXVortexGas}
\prob_{U_x}(\uc_\xc)
=
\frac{1}{2\pi}
\int_{-\infty}^{\infty}
\exp(-\prodrate \gamma(k))\, \ee^{-\imi k\uc_\xc}
\dint k,
\end{equation}
where we have dropped the superscript $\numV \rightarrow \infty$ on~$U_x$.

Since $1-\sinc(x) \sim \tfrac16 x^2$ as $x \rightarrow 0$, we have $\gamma(k) \sim \tfrac13\mathcal{E}_1 k^2$ as $k \rightarrow 0$, from which we can solve for an approximate velocity distribution, valid as $\volfrac = \prodrate R_0^5 / \nu \gg 1$: \begin{equation}
\label{approximateVelocityDistribution}
\prob_{U_x}(\uc_\xc)
\approx
\sqrt{\frac{3}{4 \pi \prodrate \mathcal{E}_1}}
\, \exp \left( - \frac{3 \uc_\xc^2}{4 \prodrate \mathcal{E}_1} \right),
\end{equation}
consistent with the central limit theorem. Then
$\langle \uc_\xc^2 \rangle
=
\frac{2}{3} \langle E \rangle
=
\frac{2}{3} \prodrate \mathcal{E}_1$,
as predicted by \eqref{eq:EE}.  Of course, in the limit of large~$\volfrac$ our linear superposition assumption breaks down, so~\eqref{approximateVelocityDistribution} is unlikely to be observed in practice.

To find an approximation of the probability density function which is valid for small $\volfrac$, where our model applies, we can use the probability distribution $\prob_{U^1}(\uc)$ from \eqref{ppdf} to find an approximation of $\gamma$ which is valid for large $k$ in the limit as $V, \tau \rightarrow \infty$:
\begin{equation}
\label{eq:gammaApprox}
\gamma(k)
=
V\tau \int_0^\infty \{ 1 - \sinc(k \uc) \}\, \prob_{U^1}(\uc)\,\dint\uc
\sim
0.1096\, \frac{\Gamma_0^{5/3} R_0^{10/3}}{\nu} |k|^{5/3} =: \frac{\hyperParam}{\prodrate} |k|^{5/3},
\end{equation}
(with $\hyperParam = 0.1096\,\prodrate (\Gamma_0 R_0^2)^{5/3} / \nu = 0.1096\, (\Gamma_0/R_0)^{5/3}\volfrac$) where we have compensated for the uniform 40\% overestimate of $\prob_{U^1}(\uc)$ by \eqref{ppdf}, as observed in Figure \ref{fig:singleVR_PDF} by decreasing the prefactor to match numerical estimates.
We can compute \eqref{probDensityVelocityXVortexGas} analytically using this $\gamma$; the result is a $\frac53$-stable distribution, expressed in terms of hypergeometric functions in Appendix~\ref{AnalyticPDF}, Eq.~\eqref{eq:vortexGasProbDistributionEstimate}. For large $\uc_x$, \eqref{eq:vortexGasProbDistributionEstimate} reduces to
\begin{equation}
\label{bigux}
\prob_{U_x}(\uc_x) \sim
\frac{1}{2\pi}\,\Gamma\!\left( \tfrac{8}{3} \right) \hyperParam\,|\uc_x|^{-8/3},
\qquad
\volfrac^{3/5} \ll \frac{\uc_x R_0}{\Gamma_0} \ll 1,
\end{equation}
while for small $\uc_x$ the core region is reasonably well approximated by a Gaussian
\begin{equation}
\label{smallux}
\prob_{U_x}(\uc_x) \sim 0.2844 \hyperParam^{-3/5} \exp \left( -\uc_x^2/3.198 \hyperParam^{6/5} \right),
\qquad
\frac{\uc_xR_0}{\Gamma_0} \ll \volfrac^{3/5}.
\end{equation}
These forms come into alignment using asymptotic matching when $\uc_x \propto \hyperParam^{3/5}$. Of particular note, we see here that the width of the core scales as $\volfrac^{3/5}$.
Comparing with \eqref{approximateVelocityDistribution}, it is clear that \eqref{bigux} and \eqref{smallux} are only valid when $\volfrac \ll 1$; that is, even though~\eqref{smallux} resembles a Gaussian distribution, it is completely different from the Gaussian~\eqref{approximateVelocityDistribution} in the large~$\volfrac$ limit.  Moreover, the tail distribution~\eqref{bigux} contributes heavily to the energy~$\mathcal{E}_1$, which therefore cannot be deduced from the width of~\eqref{smallux}.

The $-8/3$ power law in~\eqref{bigux} does not persist for arbitrarily large~$\uc_x$, and in fact one can show using an argument similar to that in Section~\ref{Single_VR_pdf} that $\prob_{U_x}(\uc_x) \propto |\uc_x|^{-5}$ as $|\uc_x| \rightarrow \infty$ due to the singular behavior of a vortex ring at $\rho = R_0$ and $\time = 0$.
Including the large $\uc$ behavior in our calculations changes the distribution from a stable distribution to a truncated stable distribution, which has finite second moment (and thus finite energy). This observation explains the seemingly inconsistent large and small $\volfrac$ approximations for $\prob_{U_x}(\uc_x)$ of a Gaussian and a stable distribution, respectively.
The transition from a truncated stable distribution to a Gaussian distribution occurs near a volume fraction where the width of the core region is on the same order of magnitude as the cutoff, which follows immediately from the Berry--Ess\'een theorem \cite{Shlesinger1995}. For further discussion of the relative contributions of the core and the tails to the energy, both with and without truncation, see Appendix \ref{Energy_contributions}.

\subsection{Comparison with numerical simulations}
\label{numeric_results}

Since a number of approximations were used to derive the distributions in the previous section, a comparison with numerical simulations is in order.  In particular, in computing \eqref{probDensityVelocityXVortexGas} we inserted a cutoff between the $-8/3$ and $-5$ power laws, and the use of \eqref{eq:gammaApprox} is not valid for small $k$.

Our numerical investigation involves a Monte Carlo integration of \eqref{probDensityVelocityXSingleVR}: we simulate the suspension by generating and evolving vortex rings uniformly in time and space in a spherical volume of radius $L = 100 R_0$ for $\time \in [0,\tau]$ with $\tau = 100 R_0^2 / \nu$ and computing the velocity at the origin. We fix the initial single-vortex circulation to be $\Gamma_0 = 100 \nu$, so all the vortices have the same initial strength. The velocity field due to individual vortices is obtained by differentiating the streamfunction \eqref{eq:IC_streamfunction} using a fourth-order-accurate finite-difference approximation.  The velocity fields of individual vortices are then superimposed linearly to generate the total velocity field. This is a reasonable approximation in the dilute regime, $\volfrac \ll 1$, when vortices stay far enough apart so that they do not significantly interact.

Because of the special functions and the oscillatory integrand,  the streamfunction $\Psi$ is prohibitively expensive to evaluate directly.   We compute it for several points on two overlapping grids and form a cubic spline interpolant to evaluate it at arbitrary points in space. One grid covers $\rho, |z| \le 20 R_0$ and $0 \le \time \le 20 R_0^2/\nu$ with $200^3$ grid points, while another grid with higher resolution covers $0.75 R_0 \le \rho \le 1.25 R_0$, $|z| \le 0.25 R_0$ and $0 \le \time \le 0.5 R_0^2 / \nu$, with $250^2 \times 100$ grid points around the initial singularity. For points outside these grids, $\Psi$ is approximated using \eqref{psiApproximation}. Since the interpolated values of $\Psi$ do not match \eqref{psiApproximation} on the boundary of the grid, a buffer region is established where $\Psi$ is represented as a convex combination of the interpolated value and \eqref{psiApproximation}; the smoothness of the transition is important in order to accurately compute the velocity. The integration required to compute $\Psi$ in \eqref{eq:IC_streamfunction} at any particular grid point is performed using a global adaptive quadrature (Matlab's integral function) with absolute and relative error tolerances $10^{-10}$ and $10^{-6}$, respectively.
A single simulation amounts to placing a random distribution of vortices, each with a random position, orientation, and age, and using the machinery above to compute the velocity at the origin at that moment.

\begin{figure}
	\centering
	\includegraphics[width=0.49\textwidth]{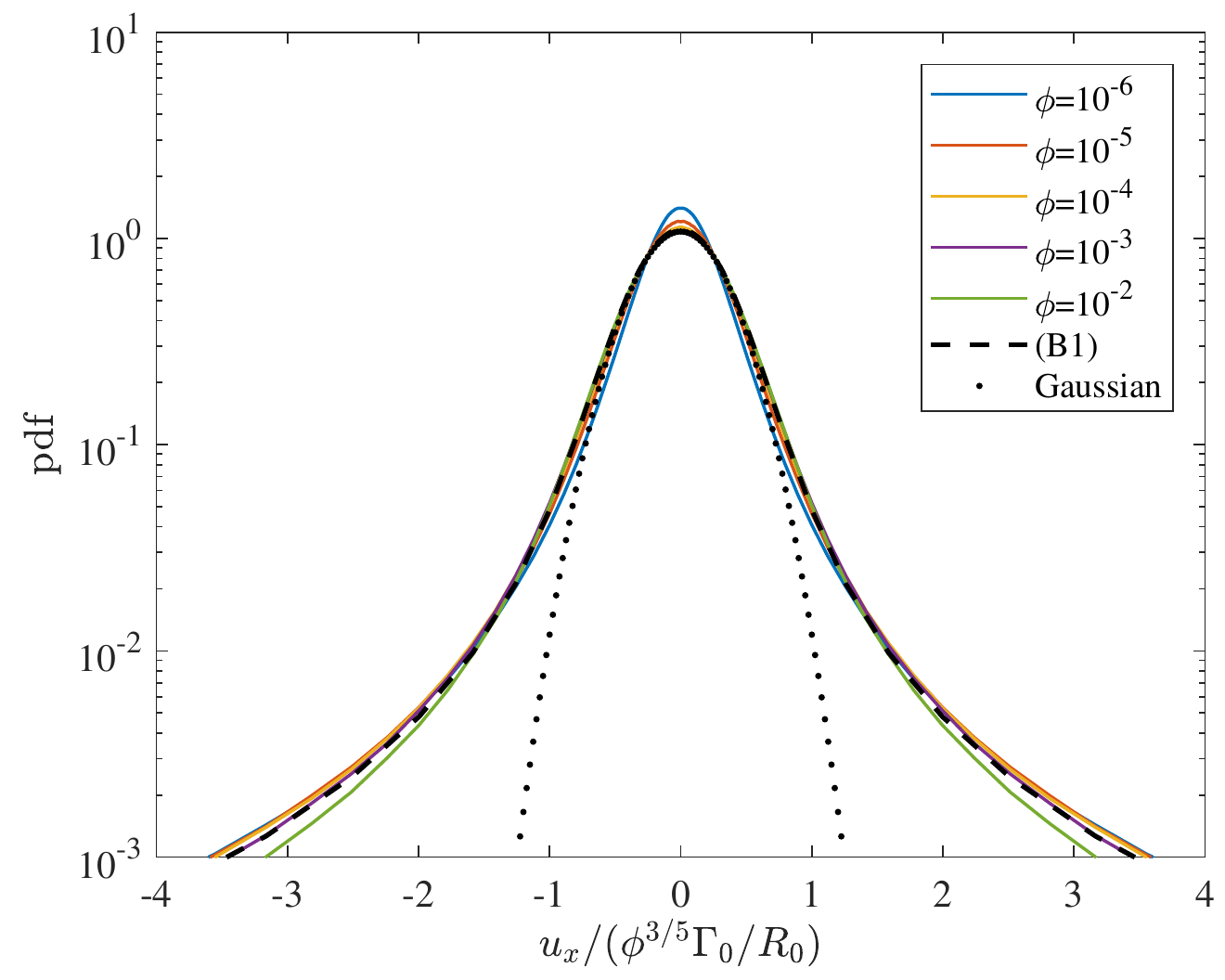}
	\caption{The probability density function for the $x$-component of velocity (normalized by $\volfrac^{3/5} \Gamma_0 / R_0$) for various $\volfrac$. We see that the core scales with $\volfrac^{3/5}$. The dashed curve is from the analytical expression \eqref{eq:vortexGasProbDistributionEstimate}, showing close agreement with the numerics. The dotted curve is a Gaussian distribution with unit standard deviation, included for reference.}
	\label{fig:normalizedVelocityPDF}
\end{figure}

\begin{figure}
	\centering
	\includegraphics[width=0.49\textwidth]{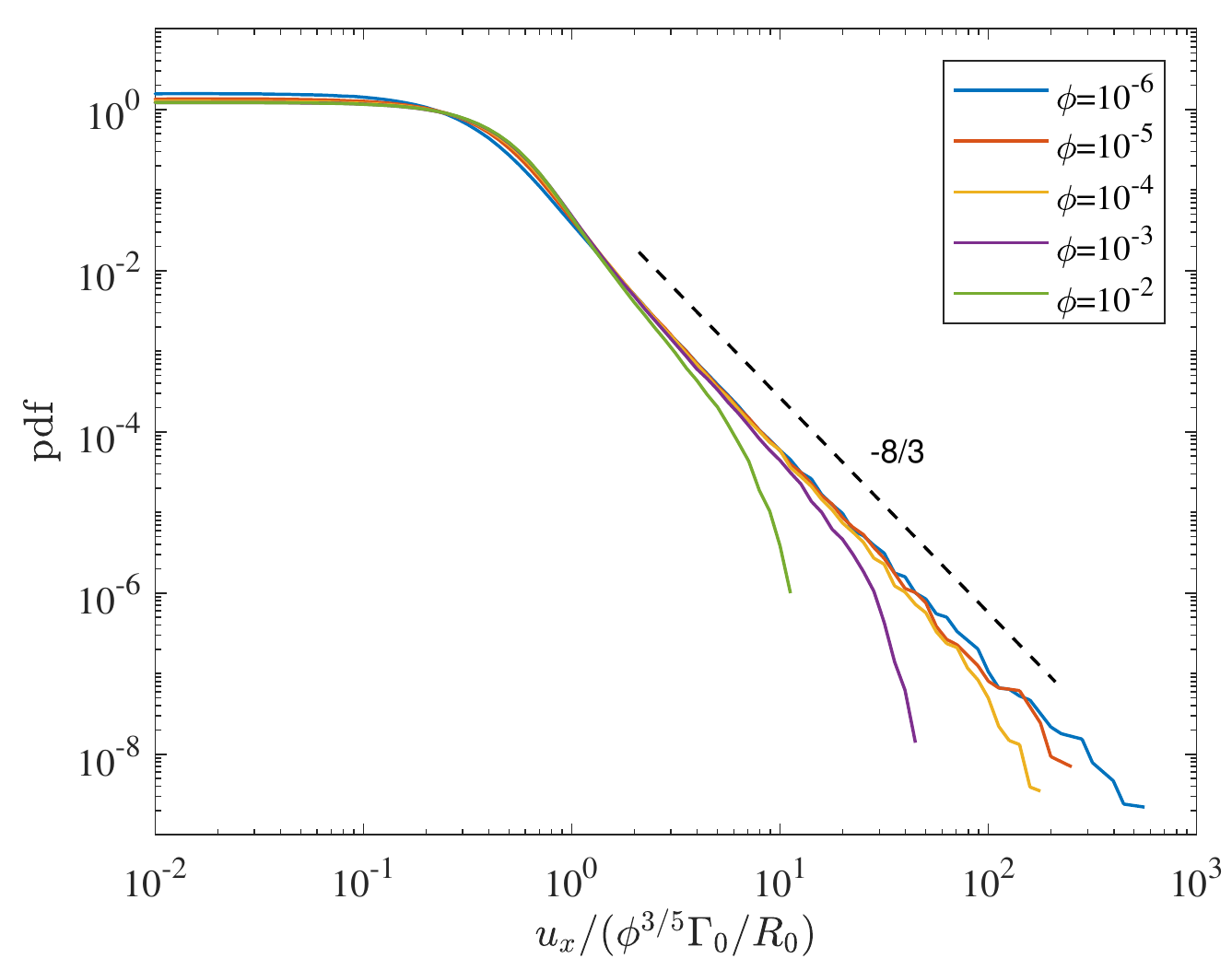}
	\caption{The same distributions as in Figure \ref{fig:normalizedVelocityPDF}, but on a log-log scale. The additional dashed line verifies the $-8/3$ power law for large (but not very large) velocities.}
	\label{fig:velocityPDF}
\end{figure}

For a given value of the effective volume fraction $\volfrac = \prodrate R_0^5 / \nu$ we run 15 million simulations on a distributed computing framework and then compute the probability density function $\prob_{U_x}(\uc_x)$ for a single component of velocity by placing the results in exponentially-sized bins. Figure \ref{fig:normalizedVelocityPDF} shows this density normalized for a selection of different $\volfrac$, along with the theoretical expression \eqref{eq:vortexGasProbDistributionEstimate} as a dashed line and a Gaussian distribution as a dotted line for reference. The numerical simulations appear to confirm the accuracy of \eqref{eq:vortexGasProbDistributionEstimate} for the entire range of $\volfrac$ considered. Note in particular the scaling of the core width by $\volfrac^{3/5}$.
Figure \ref{fig:velocityPDF} shows the same distributions on a log-log scale, with a dashed line of slope $-8/3$ included for reference. The probability density function decays as $|\uc_x|^{-8/3}$ outside the core, as predicted in \eqref{bigux}. We were unable to verify the predicted $|\uc_x|^{-5}$ power law for very large velocities due to the extreme resolution needed near the initial vortex filaments in order to properly capture the largest velocities.

\begin{figure}
	\centering
	\includegraphics[width=0.49\textwidth]{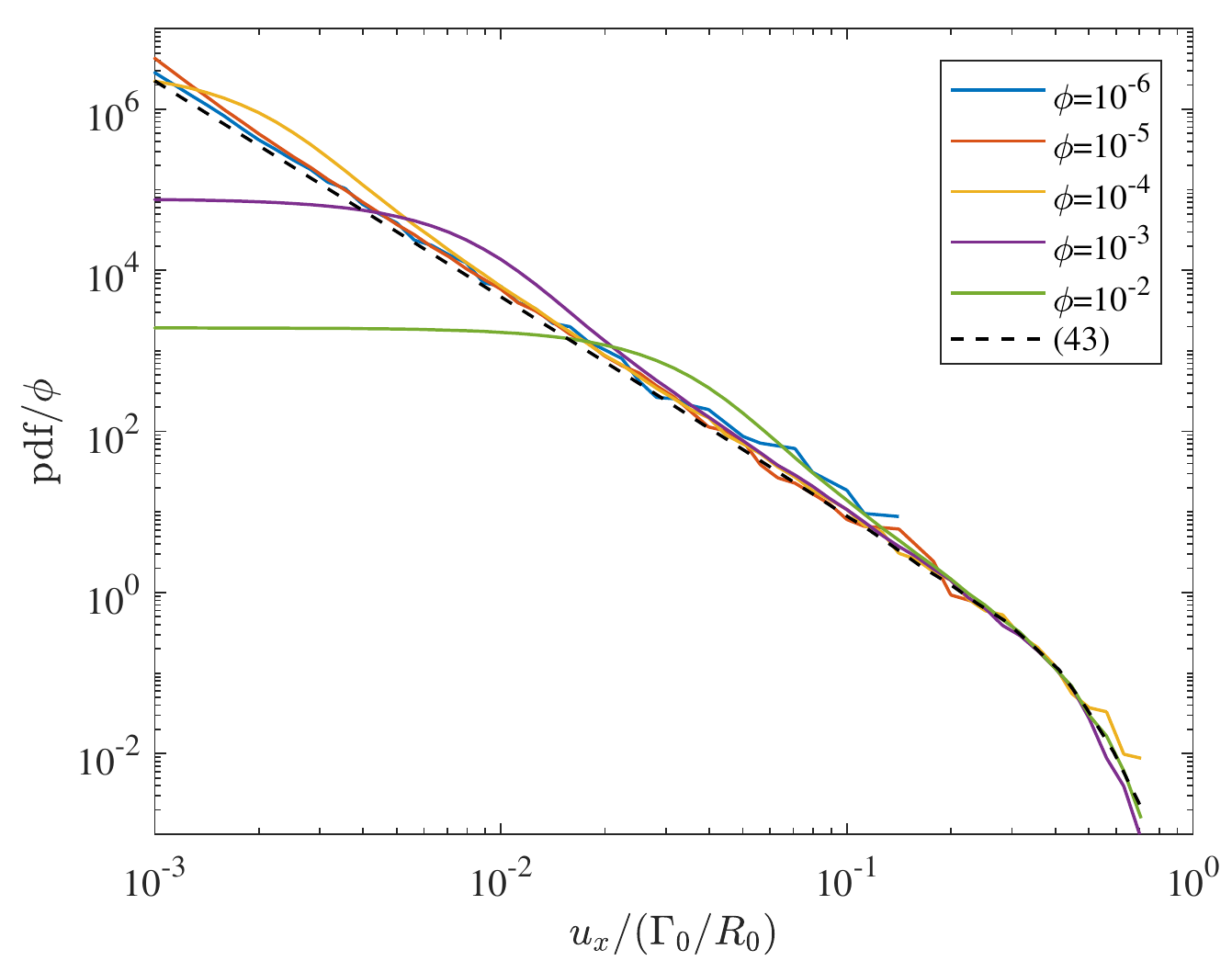}
	\caption{Plot of the (normalized) probability density function for the $x$-component of velocity divided by $\volfrac$ compared with \eqref{eq:largeDevScaling} (the dashed line), showing close agreement, except at small velocities. In particular, regardless of $\volfrac$, the distributions transition away from the $-8/3$ power law at around $\uc_x \approx 0.4 \Gamma_0 / R_0$, regardless of $\volfrac$.}
	\label{fig:compare_to_single_vr}
\end{figure}

For large enough velocities, the nearest vortex ring determines the velocity at a point, so that the many-vortex probability distribution $p_{U_x}(\uc_x)$ has the same tails as the single-vortex $p_{U_x^1}(\uc_x)$.  In particular, outside the core of the distribution we have
\begin{equation}
  \label{eq:largeDevScaling}
  \prob_{U_x}(\uc_x)
  \sim
  \volfrac\,p_{U_x^1}(\uc_x)
  =
  \volfrac \int_{|\uc_x|}^{\infty} \frac{\prob_{U^1}(\uc)}{2\uc} \,\dint\uc,
  \qquad
  \frac{\uc_xR_0}{\Gamma_0} \gg \volfrac^{3/5}\,.
\end{equation}
Figure \ref{fig:compare_to_single_vr} compares \eqref{eq:largeDevScaling} (dashed curve) with PDFs divided by $\volfrac$ for several values of~$\volfrac$. There is excellent agreement outside the core of the distribution, so typical velocities in the suspension are indeed dominated by the nearest vortex ring except in the case of small velocities.

Figures \ref{fig:normalizedVelocityPDF}--\ref{fig:compare_to_single_vr} suggest strongly that $\prob_{U_x}(\uc_x)$ is a truncated stable distribution with smooth cutoff near $\uc_x \approx \pm 0.4 \Gamma_0 / R_0$. Note that this cutoff is independent of $\volfrac$ and only depends on the transition between small and large $\uc$ asymptotics for the velocity distribution of a single vortex ring. When $\volfrac \ll 1$, the cutoff is far down the tail, so a stable distribution is a good approximation for the velocity distribution. 

\section{Robustness}
\label{Robustness}

\mathnotation{\dens}{\rho_0}

In this section we consider the flow due to an arbitrary impulsive force localized near the origin in time and space and find the same far-field behavior as in the previous section. Thus, the analysis from the last section (except for the large-velocity $|\uc_\xc|^{-5}$ tails, which are specific to the vortex model) is generic and carries through to more general flows.

Consider an external force density
\begin{equation}
\widetilde{\force}(\rv,\time)
=
\dens\,\force(\rv)\, \frac{1}{\Delta \time}\, g(\time/\Delta \time),
\label{eq:forceRobust}
\end{equation}
where~$\dens$ is the constant fluid density, $g(s)$ is nonnegative with unit integral and with support contained in $[0,1]$, and $\force(\rv)$ has compact support encompassing the origin. In the limit as $\Delta \time \rightarrow 0$, a classical argument (see for example \citet{Buhler2007}) shows that the nonlinear terms in the incompressible Navier--Stokes equations are negligible when considering the evolution due to this force of a fluid initially at rest.  The pressure $\tilde{\pressure}$ then satisfies a Poisson equation $\lapl \tilde{\pressure} = \grad \cdot \widetilde{\force}$ with boundary condition $\nabla \tilde{\pressure} \rightarrow 0$ as $r \rightarrow \infty$. \citet{Buhler2007} concludes that $\tilde{\pressure}$ has the same time dependence as $\widetilde{\force}$, i.e.,
\begin{equation}
\tilde{\pressure}(\rv,\time) = \dens\,\pressure(\rv)\, \frac{1}{\Delta \time}\, g(\time/\Delta \time).
\label{eq:pressureRobust}
\end{equation}
The linear momentum equation can be integrated over $\time \in [0,\Delta \time]$, at the end of which
\begin{equation}
\label{eq:post_impulse}
\vv(\rv,\Delta\time) + \grad\pressure(\rv)
=
\force(\rv),
\qquad
\Delta\time \rightarrow 0,
\end{equation}
where we neglected the viscous term since it is of order~$\Delta\time$ after integration. Far away from the origin, the pressure is harmonic with
\begin{equation}
\label{eq:impulse_far_field}
\pressure(\rv)
\sim
\frac{\Impv\cdot\rv}{4\pi r^3},
\qquad
r \rightarrow \infty,
\quad\text{where}\quad
\Impv = \int_{\mathbb{R}^3} \force(\rv) \dint V,
\end{equation}
so that~$\rho_0\Impv$ is the total impulsive momentum input \citep{Buhler2007}.  Substituting \eqref{eq:impulse_far_field} into \eqref{eq:post_impulse}, we find that $\vc(\rv,\Delta\time) = O(r^{-3})$ in the far field.

Taking the curl of \eqref{eq:post_impulse} gives vorticity ${\bm \omega}(\rv,\Delta \time) = \grad \times \force(\rv)$. Note that ${\bm \omega}(\rv,\Delta \time)$ has compact support contained in the support of $\force$. Assume small Reynolds number, in this section defined to be $\Rey := R_0 F / \nu$, where $F$ is a characteristic magnitude of $\force$ and $R_0$ is the radius of the smallest ball containing the support of $\force$. Then the nonlinear term in Navier--Stokes can be neglected, so the vorticity obeys a heat equation
\begin{equation}
  \frac{\partial {\bm \omega}}{\partial \time} \approx \nu \lapl {\bm \omega},
  \qquad
  \time > \Delta\time,
  \qquad
  {\bm \omega}(\rv,\Delta\time)
  = \grad \times \force(\rv).
\label{eq:vorticityUnsteadyStokes}
\end{equation}
In the limit~$\Delta\time\rightarrow0$, this has solution
\begin{equation}
  {\bm \omega}(\rv,\time)
  =
  \frac{1}{(4\pi\nu\time)^{3/2}} \int\limits_{\lvert\rv'\rvert \le R_0} [\grad \times \force(\rv^\prime)] \,\, \ee^{-\nofrac{\lvert \rv^\prime - \rv \rvert^2}{4\nu\time}} \dint V_{\rv^\prime}.
\label{eq:vorticityTimeEvolution}
\end{equation}
For $\nu\time \gg R_0 \max (R_0,\lvert\rv\rvert)$, we can expand the exponential to obtain
\begin{equation}
{\bm \omega}(\rv,\time)
=
\frac{1}{(4\pi\nu\time)^{3/2}} \int\limits_{\lvert\rv^\prime\rvert \le R_0} [\grad \times \force(\rv^\prime)] \,\, \ee^{-\nofrac{\lvert\rv\rvert^2}{4\nu\time}} \left( 1 - \frac{\lvert\rv^\prime\rvert^2 - 2 \rv \cdot \rv^\prime}{4\nu\time} + \cdots \right) \! \dint V_{\rv^\prime}.
\end{equation}
The integral of the first term in the series vanishes; the next order term gives the asymptotic behavior of the vorticity:
\begin{equation}
{\bm \omega}(\rv,\time)
\sim
\frac{\pi}{(4\pi\nu\time)^{5/2}}  \, \ee^{-\nofrac{\lvert\rv\rvert^2}{4\nu\time}}  \int\limits_{\lvert\rv^\prime\rvert \le R_0} [\grad \times \force(\rv^\prime)] \left( 2 \rv \cdot \rv^\prime - \lvert\rv^\prime\rvert^2 \right) \dint V_{\rv^\prime}, \qquad \nu\time \gg R_0 \max (R_0,\lvert\rv\rvert).
\end{equation}
An integration by parts simplifies the expression:
\begin{align}
{\bm \omega}(\rv,\time)
&=
\frac{2\pi}{(4\pi\nu\time)^{5/2}}  \, \ee^{-\nofrac{\lvert\rv\rvert^2}{4\nu\time}}  \int\limits_{\lvert\rv^\prime\rvert \le R_0} \force(\rv^\prime) \times (\rv - \rv^\prime) \dint V_{\rv^\prime} \nonumber \\
&=
\frac{2\pi}{(4\pi\nu\time)^{5/2}}  \, \ee^{-\nofrac{\lvert\rv\rvert^2}{4\nu\time}}  \left( \Impv \times \rv - {\bm J} \right),
\end{align}
where
\begin{equation}
{\bm J} := \int\limits_{\lvert \rv^\prime \rvert \le R_0} \force(\rv^\prime) \times \rv^\prime \dint V_{\rv^\prime}.
\end{equation}

The corresponding velocity field can be found via the Biot--Savart law:
\begin{equation}
\vv(\rv,\time)
=
\frac{1}{2 (4\pi\nu\time)^{5/2}} \int\limits_{\mathbb{R}^3} \frac{(\Impv \times \rv_0 - {\bm J}) \times (\rv - \rv_0)}{\lvert \rv - \rv_0 \rvert^3} \,\,\ee^{-\nofrac{\lvert \rv_0 \rvert^2}{4\nu\time}} \dint V_{\rv_0}.
\label{eq:velocityDueToImpulse}
\end{equation}
As $t \rightarrow \infty$, a vanishingly small error is introduced replacing $\rv_0$ by $\rv-\rv_0$ in the exponential. Then
\begin{align}
  \vv(\rv,\time)
  &\sim
  \frac{1}{2(4\pi\nu\time)^{5/2}} \int\limits_{\mathbb{R}^3} \frac{(\Impv \times (\rv_0 - \rv) + (\Impv \times \rv - {\bm J})) \times (\rv - \rv_0)}{\lvert \rv - \rv_0 \rvert^3} \,\, \ee^{-\nofrac{\lvert \rv - \rv_0 \rvert^2}{4\nu\time}} \dint V_{\rv_0} \nonumber\\
  &= \frac{\Impv}{12 (\pi\nu\time)^{3/2}}\,.
\end{align}
This matches \eqref{BasicVelApprox} ($\xi \gtrsim 1$) for $\Impv = \pi \Gamma_0 R_0^2\, \zuv$, the hydrodynamic impulse for the model vortex ring. For this value of impulse, the velocity field obtained by substituting \eqref{eq:impulse_far_field} into \eqref{eq:post_impulse} also matches the $\xi \lesssim 1$ case.

We see that in the limit as $r \rightarrow \infty$, the velocity decays as $O(r^{-3})$, and for any fixed location, the velocity decays as $O(t^{-3/2})$ as $t \rightarrow \infty$. The transition between these two regimes occurs along the same viscous front as we have already analyzed for the vortex ring. Indeed, \eqref{BasicVelApprox} is a good approximation for the velocity away from the impulse for any flow due to a localized impulsive force. Therefore, all our analysis from the previous section carries through and so \eqref{eq:vortexGasProbDistributionEstimate} and \eqref{bigux}--\eqref{smallux} give an approximation of the velocity distribution for a volume of fluid containing any swimmers that exert force in short bursts, such as for instance copepods \cite{Jiang2007,Jiang2011b}.

\section{Discussion}
\label{discussion}

We analyzed the flow field of a model viscous vortex ring and found that for a flow which is initially a vortex filament, the absolute moments of velocity $M_n$ are finite only for $\tfrac53 < n < 4$. Consistent with this observation, the density function of the magnitude of velocity is asymptotic to $\uc^{-8/3}$ for small velocities, and to $\uc^{-5}$ for large velocities. The former power law is due to the long-time diffusion of vorticity as the vortex ring expands, while the latter is due to the initial diffusion of vorticity away from the vortex filament immediately after its formation. While the large $\uc$ distribution will depend heavily on the exact model used, the $\uc^{-8/3}$ power law for small velocities is robust in the sense that any flow brought about by an initial impulse will produce a distribution with the same power law.

We have constructed a model suspension of viscous vortex rings with convenient analytic properties by superimposing the flow fields for individual vortex rings positioned and oriented randomly throughout space and time. The velocity fluctuations were shown both analytically and numerically to fit a truncated stable distribution with tails decaying as $\uc^{-8/3}$.  This distribution has core width proportional to $\volfrac^{3/5}$ but energy proportional to $\volfrac$, the vortex volume fraction, so that most of the energy comes from the tail of the distribution (associated with large velocities). Points in space corresponding to the distribution's tail are only influenced by the nearest vortex ring, so interactions between vortices play a negligible role. However, with increasing volume fraction $\volfrac$, the dominant contribution begins to come from the core region encompassing the far-field velocity of many not-so-distant vortices.

Our work extends efforts to understand the velocity fluctuations produced by swimmers at low Reynolds numbers to intermediate values.  We expect the model to provide a good approximation for the flow fields associated with a variety of jellyfish species in a physically-realistic regime of the Reynolds number ($60 \lesssim \Rey \lesssim 2160$) \cite{Olesen1994}, particularly in light of the robustness of the flow structure to perturbations of the initial impulse. Even among jellyfish, however, different types of flow fields are generated by different species: elongated jellyfish such as {\it Nemopsis bachei} generate a streak of vortex rings for efficient swimming \cite{Costello2002,Costello2006} while more bulbous species like {\it Aurelia aurita} generate dual starting and stopping vortex rings (during power and recovery strokes) in the wake of the bell in a slower, axisymmetric-paddling locomotion \cite{Costello2002,Gharib2005,Miller2017}. The extent to which the distribution derived here remains appropriate for describing such systems, and related non-motile systems like pulsing corals \cite{Miller2017b}, remains an open question for future exploration.

\appendix

\section{The probability density function for single vortex ring}
\label{Single_VR_pdf_Computation}

For $\xi \le 1$, the velocity is only a function of time ($\time = (\Gamma_0 R_0^2 / 12 \uc \sqrt{\pi} \nu^{3/2})^{2/3}$), so
\begin{align}
  \label{pdf_velocity_late_time_single_vr}
\int\limits_{\vc(\rv,\time) = \uc,\, \xi \le 1} \frac{\dint S_{\rv,\time}}{| \grad_{(\rv,\time)} v(\rv,\time) |}
&=
\tfrac43 \pi (4 \nu \time)^{3/2}
\left(
\frac{\Gamma_0 R_0^2}{8 \sqrt{\pi} \nu^{3/2} \time^{5/2}}
\right)^{-1}
{\bigg|}_{t = (\Gamma_0 R_0^2 / 12 \uc \sqrt{\pi} \nu^{3/2})^{2/3}} \nonumber\\
&=
\frac{2^{8/3} \pi^{1/6}}{3^{11/3}}
\frac{\Gamma_0^{5/3} R_0^{10/3}}{\nu}\,\uc^{-8/3}.
\end{align}
The integral for $\xi \ge 1$ is somewhat more complicated. From \eqref{BasicVelApprox}, we see that
\begin{equation}
\label{v_xi_ge_1}
\vc =
\frac{\Gamma_0 R_0^2}{r^3} \frac{\sqrt{4 - 3 \cos^2 \azAng}}{4} =: \frac{\Gamma_0 R_0^2}{r^3} \angFunc,
\end{equation}
where $\azAng$ is the angle from the positive $z$-axis. When the velocity is $\uc$, $r = (\Gamma_0 R_0^2 \angFunc / \uc)^{1/3}$. Then
\begin{multline}
\label{pdf_velocity_far_out_single_vr}
\int\limits_{\vc(\rv,\time) = \uc, \xi \ge 1} \frac{\dint S_{\rv,\time}}{| \grad_{(\rv,\time)} v(\rv,\time) |} \\
=
\int_0^\pi
\int_0^{r_\uc(\azAng)^2/4\nu}
\left( \frac{\uc}{r_\uc(\azAng)} \sqrt{9 + \cfrac{\dAngFunc^2}{\angFunc^2}} \right)^{-1}
2 \pi r_\uc(\azAng) \sqrt{r_\uc(\azAng)^2 + r_\uc^\prime(\azAng)^2} \sin \azAng
\,\dint \time \,\dint \azAng,
\end{multline}
where we have parameterized our surface in $\theta, \azAng, \time$ and performed the integral over $\theta$. The integral in \eqref{pdf_velocity_far_out_single_vr} can be computed analytically:
\begin{align}
\label{pdf_velocity_xi_ge_1_single_vr}
\frac{2\pi}{\nu \uc}
\int_0^\pi \frac{r_\uc(\azAng)^4 \sqrt{r_\uc(\azAng)^2 + r_\uc^\prime(\azAng)^2}}{\sqrt{9\angFunc^2+\dAngFunc^2}} \angFunc \sin \azAng \,\dint \azAng
&=
\frac{2\pi}{3} \frac{\Gamma_0^{5/3} R_0^{10/3}}{\nu}\,\uc^{-8/3}
\int_0^\pi \angFunc^{13/3} \sin \azAng \,\dint \azAng\nonumber\\
&=
0.05909 \frac{\Gamma_0^{5/3} R_0^{10/3}}{\nu}\,\uc^{-8/3}.
\end{align}

\section{Analytic probability density function}
\label{AnalyticPDF}

We can compute \eqref{probDensityVelocityXVortexGas} analytically with $\gamma$ taken from \eqref{eq:gammaApprox}:
\begin{align}
\label{eq:vortexGasProbDistributionEstimate}
\begin{split}
p_{U_x}(\uc_x)
&=
\frac{3^{11/10} \Gamma(\tfrac{8}{15}) \Gamma(\tfrac{13}{15}) \Gamma(\tfrac{1}{5})}{10 \pi^2 \hyperParam^{3/5}}
\!\,{}_4F_7\!\left( \tfrac{4}{15}, \tfrac{13}{30}, \tfrac{23}{30}, \tfrac{14}{15}; \tfrac{1}{5}, \tfrac{3}{10}, \tfrac{2}{5}, \tfrac{1}{2}, \tfrac{7}{10}, \tfrac{4}{5}, \tfrac{9}{10}; -\hyperArg \right) \\
&\hspace{0.25in} -
\frac{3^{23/10} \uc_x^2 \Gamma(\tfrac{1}{5}) \Gamma(\tfrac{14}{15}) \Gamma(\tfrac{19}{15})}{2^{11/5} 5^2 \pi^{3/2} \hyperParam^{9/5} \Gamma(\tfrac{11}{10})}
\,{}_4F_7\!\left( \tfrac{7}{15}, \tfrac{19}{30}, \tfrac{29}{30}, \tfrac{17}{15}; \tfrac{2}{5}, \tfrac{1}{2}, \tfrac{3}{5}, \tfrac{7}{10}, \tfrac{9}{10}, \tfrac{11}{10}, \tfrac{6}{5}; -\hyperArg \right) \\
&\hspace{0.25in} -
\frac{3^{27/10} \uc_x^6 \Gamma(\tfrac{26}{15}) \Gamma(\tfrac{31}{15}) \Gamma(\tfrac{12}{5})}{1120 \pi^2 \hyperParam^{21/5}}
\,{}_4F_7\!\left( \tfrac{13}{15}, \tfrac{31}{30}, \tfrac{41}{30}, \tfrac{23}{15}; \tfrac{4}{5}, \tfrac{9}{10}, \tfrac{11}{10}, \tfrac{13}{10}, \tfrac{7}{5}, \tfrac{3}{2}, \tfrac{8}{5}; -\hyperArg \right) \\
&\hspace{0.25in} +
\frac{3^{29/10} \uc_x^8 \Gamma(\tfrac{32}{15}) \Gamma(\tfrac{37}{15}) \Gamma(\tfrac{14}{5})}{8960 \pi^2 \hyperParam^{27/5}}
\,{}_4F_7\!\left( \tfrac{16}{15}, \tfrac{37}{30}, \tfrac{47}{30}, \tfrac{26}{15}; \tfrac{11}{10}, \tfrac{6}{5}, \tfrac{13}{10}, \tfrac{3}{2}, \tfrac{8}{5}, \tfrac{17}{10}, \tfrac{9}{5}; -\hyperArg \right) \\
&\hspace{0.25in} +
\frac{\uc_x^4}{20 \pi \hyperParam^3}
\,{}_5F_8\!\left( \tfrac{2}{3}, \tfrac{5}{6}, 1, \tfrac{7}{6}, \tfrac{4}{3}; \tfrac{3}{5}, \tfrac{7}{10}, \tfrac{4}{5}, \tfrac{9}{10}, \tfrac{11}{10}, \tfrac{6}{5}, \tfrac{13}{10}, \tfrac{7}{5}; -\hyperArg \right),
\end{split}
\end{align}
where $\hyperParam = 0.1096\, (\Gamma_0/R_0)^{5/3}\volfrac$ from \eqref{eq:gammaApprox} and
\begin{equation}
\label{hypergeomB}
\hyperArg := \frac{3^6}{2^4 5^{10}} \frac{\uc_x^{10}}{\hyperParam^6}.
\end{equation}
Caution should be taken when using this expression for numerical purposes since there is a large and increasing amount of cancellation between the terms of \eqref{eq:vortexGasProbDistributionEstimate} as $\uc_x$ increases.

\section{Energy contributions from sections of the PDF}
\label{Energy_contributions}

The expected energy of the suspension of vortices is
\begin{equation}
\label{eq:EnergyFromPDF}
\langle E \rangle = \tfrac32 \int_{-\infty}^{\infty} \uc_x^2\, \prob_{U_x}\!(\uc_x) \dint \uc_x.
\end{equation}
Equations~\eqref{bigux}--\eqref{smallux} cannot be used by themselves to approximate the energy, since this results in divergence in the expression above, so the $|\uc_x|^{-5}$ tails for the largest velocities must be included in order to obtain a convergent integral.

Using \eqref{bigux}--\eqref{smallux} to determine the behavior of the inner and middle regions, we find that
\begin{equation}
\label{threeRegionUx}
\prob_{U_x}(\uc_x) \approx \begin{cases}
0.2844 \hyperParam^{-3/5} \exp \left( -\uc_x^2/3.198 \hyperParam^{6/5} \right) & |\uc_x| \le 3.260 \hyperParam^{3/5}, \\
0.2395 \hyperParam |\uc_x|^{-8/3} & 3.260 \hyperParam^{3/5} \le |\uc_x| \le \cutoff, \\
0.2395 \hyperParam \cutoff^{7/3} |\uc_x|^{-5} & |\uc_x| \ge \cutoff,
\end{cases}
\end{equation}
where $\cutoff \approx 0.4 \Gamma_0 / R_0$, as in Figure \ref{fig:compare_to_single_vr}, and $\hyperParam = 0.1096\, (\Gamma_0/R_0)^{5/3}\volfrac$ (from \eqref{eq:gammaApprox}). A comparison to \eqref{eq:vortexGasProbDistributionEstimate} suggests that \eqref{threeRegionUx} somewhat underestimates $\prob_{U_x}\!(\uc_x)$ around the transition at $|\uc_x| = 3.260 \hyperParam^{3/5}$.

Let $\langle E_{\text{C}}\rangle, \langle E_{-8/3}\rangle$, and $\langle E_{-5}\rangle$ be the portions of the energy using the approximations of $\prob_{U_x}(\uc_x)$ in the core (C), middle ($-8/3$), and outer ($-5$) regions in \eqref{threeRegionUx} with the appropriate bounds, so that $\langle E \rangle = \langle E_{\text{C}} \rangle+ \langle E_{-8/3} \rangle+ \langle E_{-5}\rangle$. We find the contributions
\begin{subequations}
\begin{align}
\langle E_{\text{C}} \rangle&= 1.980 \hyperParam^{6/5}, \\
\langle E_{-8/3} \rangle&= -3.196 \hyperParam^{6/5} + 2.156 \hyperParam \cutoff^{1/3}, \\
\langle E_{-5} \rangle&= 0.3593 \hyperParam \cutoff^{1/3}.
\end{align}
\end{subequations}
Without the underestimate of $\prob_{U_x}\!(\uc_x)$ in the transition between the core and middle regions, the $\hyperParam^{-6/5}$ terms above should cancel exactly (since the energy is known to scale with $\volfrac$ and $\hyperParam$ is linear in $\volfrac$), which we verified using \eqref{eq:vortexGasProbDistributionEstimate} directly and integrating numerically. Thus, a rough estimate of the energy is $\langle E \rangle \approx 2.515 \hyperParam \cutoff^{1/3} = 0.2031  (\Gamma_0/R_0)^{2}\volfrac$, a slight overestimate of the exact expression in \eqref{eq:EE}. Hence, we see that the greatest contribution to the energy comes from the middle region of the distribution for small $\volfrac$.  As $\volfrac$ increases, the largest contribution begins to come from the core region, which encompasses the far-field velocity of the vortices.


\let\r\rsave 
\bibliography{../vortices}

\end{document}